\documentclass{article}

\usepackage{arxiv}

\usepackage[utf8]{inputenc} 
\usepackage[T1]{fontenc} 
\usepackage{url} 
\usepackage{booktabs} 
\usepackage{graphicx}
\usepackage{natbib}
\usepackage{doi}
\usepackage{amsfonts,amssymb,amsthm,mathtools}
\usepackage{algpseudocode,algorithm}
\usepackage{float}
\usepackage{array}
\usepackage{ragged2e}
\usepackage{tikz}
\usetikzlibrary{arrows,intersections,decorations,plotmarks,math}
\usepackage{pgfplots}
\definecolor{COL1}{RGB}{94,60,153}
\definecolor{COL2}{RGB}{241,163,64}
\newtheorem{clm}{Claim}
\newtheorem{assump}{Assumption}
\DeclareMathOperator*{\argmin}{argmin}
\newcolumntype{P}[1]{>{\RaggedRight\hspace{0pt}}p{#1}}

\title{Storage placement policy for minimizing frequency deviation: A combinatorial optimization approach}

\date{} 					

\author{\href{https://orcid.org/0000-0002-8003-4248}{\includegraphics[scale=0.06]{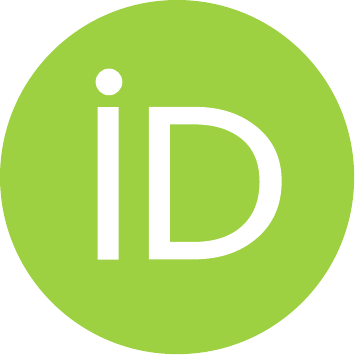}\hspace{1mm}Ram Machlev} \\
The Andrew and Erna Viterbi Faculty of Electrical Engineering \\
Technion---Israel Institute of Technology \\
Haifa 3200003, Israel \\
\texttt{ramm@campus.technion.ac.il} \\
\And
\href{https://orcid.org/0000-0001-6907-7952}{\includegraphics[scale=0.06]{orcid.pdf}\hspace{1mm}Nilanjan R.~Chowdhury} \\
The Andrew and Erna Viterbi Faculty of Electrical Engineering \\
Technion---Israel Institute of Technology \\
Haifa 3200003, Israel \\
\texttt{nilanjan2008@gmail.com} \\
\And
\href{https://orcid.org/0000-0000-0000-0000}{\includegraphics[scale=0.06]{orcid.pdf}\hspace{1mm}Juri Belikov} \\
Department of Software Science \\
Tallinn University of Technology \\
Akadeemia tee 15a, 12618 Tallinn, Estonia \\
\texttt{juri.belikov@taltech.ee} \\
\And
\href{https://orcid.org/0000-0001-9775-1406}{\includegraphics[scale=0.06]{orcid.pdf}\hspace{1mm}Yoash Levron} \\
The Andrew and Erna Viterbi Faculty of Electrical Engineering \\
Technion---Israel Institute of Technology \\
Haifa 3200003, Israel \\
\texttt{yoashl@ee.technion.ac.il} \\
}

\renewcommand{\shorttitle}{Storage placement policy for minimizing frequency deviation}

\hypersetup{
pdftitle={Storage placement policy for minimizing frequency deviation: A combinatorial optimization approach},
pdfsubject={q-bio.NC, q-bio.QM},
pdfauthor={Ram Machlev, Nilanjan R.~Chowdhury, Yoash Levron, Juri Belikov},
pdfkeywords={Distributed energy storage, Grid supporting inverters, Frequency stability, Droop control, Combinatorial optimization, Cross-entropy method},
}

\begin{document}
\maketitle

\begin{abstract}
As the share of renewable sources is increasing the need for multiple storage units appropriately sized and located is essential to achieve better inertial response. This work focuses on the question of ``how to distribute constant number of storage units in the gird under transient events such that the inertial response of the maximum frequency deviation is minimized?''. To answer this question, we provide a comprehensive modeling framework for energy storage units placement and size for frequency stability under spatial effects. The distributed storage units are modeled as grid supporting inverters and the total storage capacity in the grid is bounded based on the allowed steady-state frequency deviation after disturbances. The problem of finding the optimal distributions can be considered as combinatorial problem which consists of high dimensional solutions. In this light, we develop two numeric approaches based on Brute-force search and adaptation of the Cross-entropy method for finding the best distribution and examined it on a case study of the future Israeli grid. The results on the case study provide a new insight---the storage units should be placed around the area of the disturbances, including in sites with high inertia in accordance to the network topology.
\end{abstract}

\keywords{Distributed energy storage \and Grid supporting inverters \and Frequency stability \and Droop control \and Combinatorial optimization \and Cross-entropy method}

\section{Introduction}
The share of energy generated by renewable sources in the European Union has reached to the set target of 20\% of the total produced energy by 2020. Moreover, the call for 100\% renewable energy production worldwide in 2050 is gaining widespread support \cite{General_ReN2050}. Nonetheless, integration of renewable energy sources in existing power grids creates many challenges. One major challenge for integration of renewable energy sources in modern power systems is frequency stability. As the share of renewable sources in the grid is increasing and conventional power plants are being disconnected, the inertia within the grid is slowly being reduced. This may jeopardize the grid stability and its overall dynamic behavior \cite{General_lowInertiaSystems2018, Inertia1,Inertia2,inertia3}. One method to deal with this challenge is to install distributed fast-reacting energy storage units along the grid which absorb and discharge energy when the system frequency is not equal to its nominal value. The application of storage devices for frequency regulation has been identified as one of the applications with the highest value for storage technology \cite{General_EssApplication2006}.

An essential question is where to locate storage devices, and how to organize them on a large scale \cite{General_ESmanage2017}. Two leading concepts are the decentralized approach, which calls for numerous distributed storage units, and the centralized approach, in which relatively large storage devices are located in key points within the grid \cite{General_challengES2008}. The importance of choosing the location of storage units is mainly emphasized during the first few seconds after transient events, where generator frequencies are not equal and hence the frequency changes across the system in different locations \cite{General_FreqDiv}. A common measure for the frequency in a power system is the center of inertia frequency, i.e., the weighted average of synchronous generator rotor speeds. However, since this measure does not capture spatial effects and mainly relevant for steady-state, it may not be useful for locating storage devices along the network. Thus, when considering the spatial effects a complex question is where to locate energy storage devices with optimal size \cite{SizeLocation_reviewDistribut2019}, i.e how to consider both the location and the size of storage systems for inertial response. Due to its complexity, this question is still under study and several latest papers explore the optimal location and size for stabilizing the frequency during a contingency, such as a failure of a large synchronous generator. In work \cite{SizeLocation_2_Mexican2018} the transmission system bus with the largest frequency variation is identified, and is used as an index for energy storage placement. In addition, the sizing of the storage device is formulated as a constrained optimization problem, which is solved using a heuristic algorithm. In work \cite{SizeLocation_freqConstrains_2019} the energy storage location and size is chosen such that system frequency requirements are met during a contingency. These two studies assume that the frequency is equal throughout the grid and only one storage unit is available. Another example is \cite{SizeLocation_1_2016} where storage devices are placed at buses in which the angle variation during a contingency is highest. However, other buses in the grid are not taken into account, and the inertial response is not considered.

As concluded in \cite{essFreqResReviewFederico}, studies that involve energy storage for inertial response must not consider the frequency as uniform across the grid and should use multiple storage devices appropriately sized and located in order to achieve better and accurate frequencies regulation performance in large power systems. In accordance, during the last few years some studies which explore the problem of inertia allocation for stability considered varying frequency at different locations in the system. For example, in \cite{InertiaEffectOnFreqTrans} the grid is modeled by linear swing equations and the optimization criteria for placing grid-following virtual inertia is chosen based on damping or droop coefficients and transient overshoots while ensuring admissible transient behavior after a large disturbance. Other examples can be found in \cite{Location_ETH_PlaceVirtualInertia2017,Location_ETH2_AddVirtualInertia2017} which suggest a linear model of virtual inertia devices that modeled as local feedback control loops that connect the frequency and power injection at the terminals of a converter. In \cite{Location_ETH_PlaceVirtualInertia2017} the problem of inertia allocation is explored through the amplification of stochastic or impulsive disturbances via $\mathcal{H}_2$ performance metric. In \cite{Location_ETH2_AddVirtualInertia2017} same performance metric is used to explore the placement of virtual inertia for increasing the resilience of low-inertia power systems. Another study is \cite{Location_ETH3_PlaceInverters2019} which develops nonlinear model of converter-based virtual inertia devices that capture the key dynamic characteristics of phase-locked loops used in grid-following virtual inertia devices and of grid-forming controls. An optimization problem is formulated to optimize the parameters and location of these devices in a power system to increase its resilience. Also, work \cite{Location_FlywheelChile2017} identifies prospective location to install dedicated model of flywheel energy storage plant based on $dq0$ dynamic. An analysis to identify what are the best locations to install the plant is suggested. Lately, in paper \cite{optimalESSplaceEco2020}, a new framework is proposed, which considers the battery storage system features into the optimal placement formulation to enhance frequency response with minimum cost.

The works above formulate the problem of storage or inertia devices allocation with various objective functions, however known of them considered the inertial response for the maximum frequency deviation as the main objective when the frequency varies across the network. In light of this gap, the main contribution of this paper is to develop a numeric approach based on combinatorial optimization which allows to answer the question of ``how to distribute constant number of storage units in the gird under transient events such that the inertial response for the maximum frequency deviation is minimized?''. We focus on a model that handles distribution of storage units in a large scale power system as combinatorial problem which consists of high dimensional solutions. This work suggests a time-varying phasor model with energy distributed storage devices connected to the network using grid-supporting inverters based on droop control mechanism. In this model the total storage capacity is bounded based on the allowed steady-state frequency deviation after disturbances. Two numeric approaches are formulated using the suggested model and examined on a case study of the future Israeli grid. While the first approach, based on brute-force search, reach to global optimal solution the second approach, an adaptation of the cross-entropy method, has lower computational complexity and it may reach to optimal solution. A comprehensive analysis accompanied by comparison to case which not consider spatial effects is presented. Our numeric results conclude that the model expectation regarding size and locations of storage devices are aligned to conclusions of other state-of-the-art works. Furthermore, while other works suggest that the best locations are those located in areas with low inertia density \cite{Location_FlywheelChile2017} and that the locations of the disturbance and storage effect the resilience of the grid more than the total inertia \cite{Location_ETH_PlaceVirtualInertia2017}, this work also concludes that the storage units should be placed around the area of the disturbances, including in sites with high inertia in accordance to the network topology. The suggested approaches can provide guidelines for choosing the best locations and size of distributed storage units for frequency stability.

This paper unfolds as follows: Section~\ref{sec:models} models the overall power system network and formulates the key problem of this article. Solutions to the problem stated in Section~\ref{sec:models} are presented in Section~\ref{sec:numericAppr}. Section~\ref{sec:caseStudy} performs a series of numerical experiments on the future Israel electricity grid to verify our theoretical contributions, while Section~\ref{sec:sum} concludes this paper.\\

\textbf{Notations:} We define $\mathbb{R}$ and $\mathbb{Z}_{+}$ as the set of real numbers and positive integers, while $\mathbb{R}_{\geq 0}$ (resp.~$\mathbb{R}_{> 0}$) denotes the set of non-negative (resp.~positive) real numbers. For a matrix $M\in\mathbb{R}^{p\times q}$, $M^{\mathsf{T}}\in\mathbb{R}^{q\times p}$ denotes its transpose, and $I$ and $\mathbf{0}$ denote identity and null matrix with appropriate dimensions. The column vector $\mathbf{1}$ describes a vector in which all the entries are $1$. Given two vectors $\mathbf{a},\mathbf{b}\in\mathbb{R}^{p}$, the inequalities $\mathbf{a}>\mathbf{b}$ (resp. $\mathbf{a}<\mathbf{b}$) are considered element-wise, i.e., $a_i>b_i$ (resp. $a_i<b_i$) for each $i\in\left\lbrace 1,2,\ldots,p\right\rbrace$. For a random variable $x$, $\mathbb{E}(x)$ denotes its expected value. For two positive integer $a,b\in\mathbb{Z}_{+}$, we define ${a \choose b}:=\frac{a!}{b!(a-b)!}$.

\section{Problem setup}\label{sec:models}
In this section we seek to develop necessary technical backgrounds to formulate the key problem of this article. Towards this end, first in Section~\ref{pnm} we describe a generic power network model and then we derive the overall network dynamics. Following this, in Section~\ref{prfo} we formulate the storage units placement problem explicitly. 

\subsection{Power network model} \label{pnm}
\begin{figure}[htbp]
\centering
\includegraphics[width=0.6\textwidth]{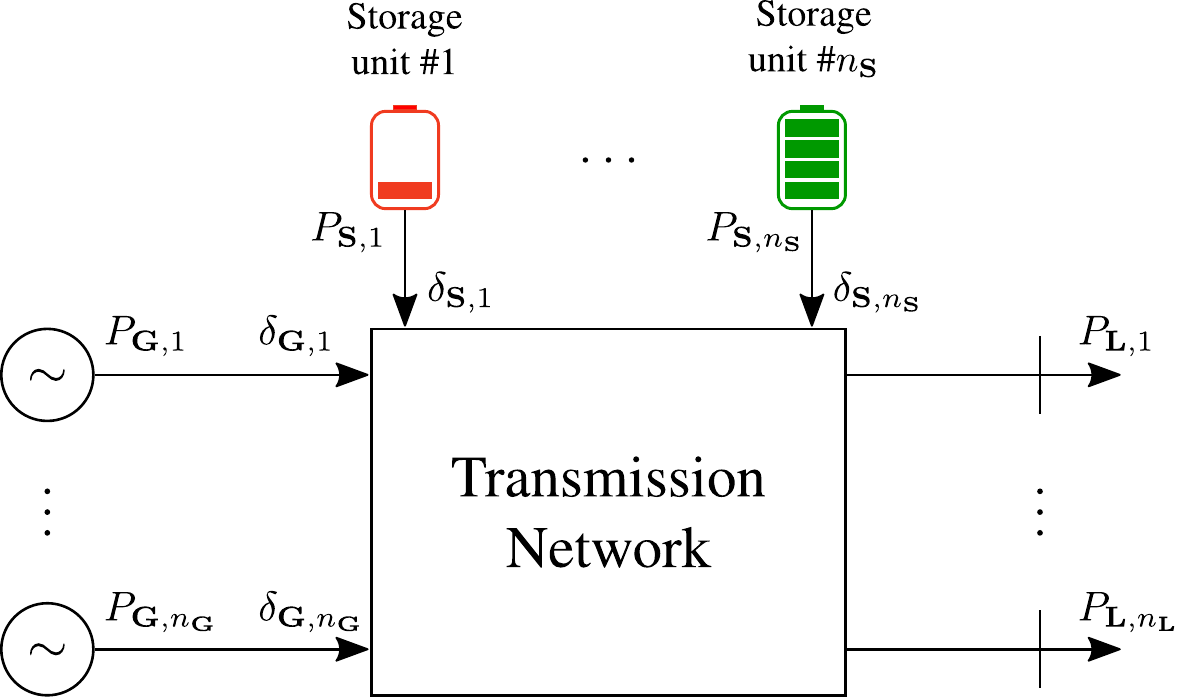}
\caption{A general power network.}
\label{fig:GridTop}
\end{figure}

We consider a general linear power system network which includes buses and transmission lines as shown in Fig.~\ref{fig:GridTop}. The overall network consists of $n$ buses among which $n_{\mathbf{G}}$ and $n_{\mathbf{L}}$ buses are connected with synchronous generators and loads (or renewable energy sources), respectively. Each bus is assumed to be either a generator or a load bus, such that it obeys $n_{\mathbf{G}}+n_{\mathbf{L}}=n$. Furthermore, we also assume if a renewable energy source is connected to a bus, then it will be considered as a negative load. 

In what follows, first in Sections~\ref{GM} and~\ref{SM} we describe the generator and storage system model considered in this work, and then in Section~\ref{TSD} we evaluate the overall power system dynamics based on these models.

\subsubsection{Generator model} \label{GM}
Let $\mathcal{N}_{\mathbf{G}}=\left\lbrace 1,2,\ldots,n_{\mathbf{G}}\right\rbrace$ be the set of synchronous generators. Then, for the $i^{th}$ generator, the dynamics of the power angle $\delta_{\mathbf{G},i}(\cdot)\in\mathbb{R}$ is defined as 
\begin{equation}\label{sg1}
\frac{\mathrm{d}}{\mathrm{d}t}\delta_{\mathbf{G},i}(t)=\omega_{\mathbf{G},i}(t)-\omega_{\mathbf{G},1}(t) \quad \forall i\in \mathcal{N}_{\mathbf{G}} \backslash \left\lbrace 1\right\rbrace.
\end{equation}
Here, $\omega_{\mathbf{G},i}(\cdot)\in\mathbb{R}$ is the frequency of the $i^{th}$ generator where $i\in \mathcal{N}_{\mathbf{G}}$, and it's dynamics is derived via the linearized \textit{swing equation} \cite{General_Kundur} as 
\begin{equation}\label{sg2}
\frac{\mathrm{d}}{\mathrm{d}t}\omega_{\mathbf{G},i}(t)=K_{i}\left(3P_{\mathbf{G},i}^{\mathrm{ref}}(t)-3P_{\mathbf{G},i}(t)-\frac{1}{D_{\mathbf{G},i}}\left(\omega_{\mathbf{G},i}(t)-\omega_0\right)\right) \quad \forall i\in \mathcal{N}_{\mathbf{G}}.
\end{equation} 
In~\eqref{sg2}, $P_{\mathbf{G},i}^{\mathrm{ref}}(\cdot),P_{\mathbf{G},i}(\cdot)\in\mathbb{R}$ and $D_{\mathbf{G},i}\in\mathbb{R}_{>0}$ are the reference power, active power per phase provided by the internal voltage source and the damping coefficient of the $i^{th}$ generator, respectively, while $\omega_0\in\mathbb{R}_{>0}$ denotes the nominal system frequency\footnote{In particular, $\omega_0$ is either $\omega_0=2\pi50$ or $\omega_0=2\pi60$ rad/s.}. The positive constant $K_i$ in~\eqref{sg2}, is defined as $K_i:=\frac{1}{J_i\omega_0}\left(\frac{p_{f,i}}{2}\right)^2$, where $p_{f,i}$ is the (positive) even number of magnetic poles of the rotor and $J_i\in\mathbb{R}_{>0}$ is the rotor moment of inertia.

\subsubsection{Storage model}\label{SM}
We consider a generalized dynamical model of the storage systems. We define $\mathcal{N}_{\mathbf{S}}=\left\lbrace 1,2,\ldots,n_{\mathbf{S}}\right\rbrace$ where $n_{\mathbf{S}}\in\mathbb{Z}_{+}$, as the set grid-connected storage devices. Then, following~\cite{chowdhury2020}, we consider the stored energy $E_{\mathbf{S},i}(\cdot)\in\mathbb{R}_{\geq 0}$ of each devices obeys the subsequent dynamics
\begin{equation}\label{st1}
\frac{\mathrm{d}}{\mathrm{d}t}E_{\mathbf{S},i}(t)= P_{\mathbf{S},i}^{\mathrm{eff}}(t) = \begin{cases}
\eta_{c,i}P_{\mathbf{S},i}(t), & P_{\mathbf{S},i}(t) > 0, \\
\eta^{-1}_{d,i}P_{\mathbf{S},i}(t), & P_{\mathbf{S},i}(t) < 0, 
\end{cases} \quad \forall i\in\mathcal{N}_{\mathbf{S}}.
\end{equation}
Here, $P_{\mathbf{S},i}^{\mathrm{eff}}(\cdot)\in\mathbb{R}$ and $P_{\mathbf{S},i}(\cdot)\in\mathbb{R}$ denote the effective power and the total power flowing into the $i^{th}$ storage device, while the constants $\eta_{c,i},\eta_{d,i}\in\left(0,1\right] $ denote its charging and discharging efficiency. Subsequent analysis assumes that each storage device is lossless, i.e., $\eta_{c,i} = \eta_{d,i} = 1$, thus, $P_{\mathbf{S},i}^{\mathrm{eff}}(t)=P_{\mathbf{S},i}(t)$ for all $t\geq 0$. Although, storage devices are capable to provide energy for a long period of time, this work predominantly focuses on the frequency stability and its impact on the inertial response during transient. Therefore, we preclude the scenarios where the storage devices are fully charged or discharged.

In this work, we consider the storage devices are deployed in conjunction with the grid supporting inverters with no voltage control which designed to provide inertia emulation and primary frequency control \cite{GridSupportInverterExample,GeneralizedDroopControlforGrid-Supporting}. Following this, the dynamical model of the $i^{th}$ storage device where $i\in\mathcal{N}_{\mathbf{S}}$ can be represented as
\begin{equation}\label{sd1}
\begin{aligned}
\frac{\mathrm{d}}{\mathrm{d}t}\delta_{\mathbf{S},i}(t)&=\omega_{\mathbf{S},i}(t)-\omega_1(t), \\
\frac{\mathrm{d}}{\mathrm{d}t}\omega_{\mathbf{S},i}(t)&= \frac{1}{\alpha_{\mathbf{S},i}}\left(3D_{\mathbf{S},i}(P_{\mathbf{S},i}^{\mathrm{ref}}(t)-P_{\mathbf{S},i}(t))-(\omega_{\mathbf{S},i}(t)-\omega_{0})\right), 
\end{aligned}
\end{equation}
where $D_{\mathbf{S},i}\in\mathbb{R}_{>0}$ is the storage device damping coefficient and the positive constant $\alpha_{\mathbf{S},i}$ defines the smoothing factor of the low pass filter. In the sequel, for each $i\in\mathcal{N}_{\mathbf{S}}$ we consider $P_{\mathbf{S},i}^{\mathrm{ref}}(t)=0$, which implies no (dis)charge during steady-state where $\omega_{\mathbf{S},i}(t)=\omega_0$.

\subsubsection{The system dynamics}\label{TSD}
In this section, we attempt to derive the dynamical model of the overall power system network. To this end, first we define $\delta(t):=\begin{bmatrix} \delta_{\mathbf{G}}^{\mathsf{T}}(t), & \delta_{\mathbf{S}}^{\mathsf{T}}(t)\end{bmatrix}^{\mathsf{T}}$ and $\omega(t):=\begin{bmatrix} \omega_{\mathbf{G}}^{\mathsf{T}}(t), & \omega_{\mathbf{S}}^{\mathsf{T}}(t)\end{bmatrix}^{\mathsf{T}}$, in which $\delta_{\mathbf{G}}(\cdot)\in\mathbb{R}^{n_{\mathbf{G}}-1}$, $\delta_{\mathbf{S}}(\cdot)\in\mathbb{R}^{n_{\mathbf{S}}}$, $\omega_{\mathbf{G}}(\cdot)\in\mathbb{R}^{n_{\mathbf{G}}}$ and $\omega_{\mathbf{S}}(\cdot)\in\mathbb{R}^{n_{\mathbf{S}}}$ are obtained as follows
\begin{align*}
\delta_{\mathbf{G}}(t)=\begin{bmatrix}
\delta_{\mathbf{G},2}\\ \vdots \\ \delta_{\mathbf{G},n_{\mathbf{G}}}
\end{bmatrix}, \quad \delta_{\mathbf{S}}(t)=\begin{bmatrix}
\delta_{\mathbf{S},1}\\ \vdots \\ \delta_{\mathbf{G},n_{\mathbf{S}}}
\end{bmatrix}, \quad \omega_{\mathbf{G}}(t)=\begin{bmatrix}
\omega_{\mathbf{G},1}\\ \vdots \\ \omega_{\mathbf{G},n_{\mathbf{G}}}
\end{bmatrix}, \quad \omega_{\mathbf{S}}(t)=\begin{bmatrix}
\omega_{\mathbf{S},1}\\ \vdots \\ \omega_{\mathbf{G},n_{\mathbf{S}}}
\end{bmatrix}.
\end{align*}
Similarly, stacking all the stored energy of the storage devices and the load power of all the load buses we obtain $\mathsf{E}_{\mathbf{S}}(t)=\begin{bmatrix}
E_{\mathbf{S},1}(t) & \ldots & E_{\mathbf{S},n_{\mathbf{S}}}(t)\end{bmatrix}^\mathsf{T}\in\mathbb{R}^{n_{\mathbf{S}}}$ and $\mathsf{P}_{\mathbf{L}}(t)=\begin{bmatrix}P_{\mathbf{L},1}(t) & \ldots & P_{\mathbf{L},n_{\mathbf{L}}}(t)\end{bmatrix}^\mathsf{T}\in\mathbb{R}^{n_{\mathbf{L}}}$. The overall reference power of the network denoted as $\mathsf{P}_{\text{ref}}(t)$, can further be evaluated as $\mathsf{P}_{\text{ref}}(t)=\begin{bmatrix} P_{\mathbf{G},\text{ref}}^{\mathsf{T}}(t), & P_{\mathbf{S},\text{ref}}^{\mathsf{T}}(t)\end{bmatrix}^{\mathsf{T}}$, where $P_{\mathbf{G},\text{ref}}(\cdot)\in\mathbb{R}^{n_{\mathbf{G}}}$ and $P_{\mathbf{S},\text{ref}}(\cdot)\in\mathbb{R}^{n_{\mathbf{S}}}$ are described subsequently
\begin{align*}
P_{\mathbf{G},\text{ref}}(t)=\begin{bmatrix}
P_{\mathbf{G},1}^{\mathrm{ref}}(t) \\ \vdots \\ P_{\mathbf{G},n_{\mathbf{G}}}^{\mathrm{ref}}(t)
\end{bmatrix}, \quad P_{\mathbf{S},\text{ref}}(t)=\begin{bmatrix}
P_{\mathbf{S},1}^{\mathrm{ref}}(t) \\ \vdots \\ P_{\mathbf{S},n_{\mathbf{S}}}^{\mathrm{ref}}(t)
\end{bmatrix}.
\end{align*} 
Now to obtain the overall power system dynamics we seek to invoke results from the DC power flow equations. Toward this end, following the DC power flow equations given in~\cite[Chapter~$12$]{das2017power}, we obtain
\begin{equation}\label{dc1}
\mathsf{P}(t)=\mathcal{G}\delta(t)+\mathcal{H}\mathsf{P}_{\mathbf{L}}(t).
\end{equation}
Here, the vector $\mathsf{P}(\cdot)\in\mathbb{R}^{n_{\mathbf{G}}+n_{\mathbf{S}}}$ is obtained by stacking the active powers of all the generators and the loads i.e. $\mathsf{P}(t)=\begin{bmatrix} P_{\mathbf{G},1} & \ldots & P_{\mathbf{G},n_{\mathbf{G}}} & P_{\mathbf{S},1} & \ldots & P_{\mathbf{S},n_{\mathbf{S}}} \end{bmatrix}^\mathsf{T}$. Furthermore, the matrices $\mathcal{G}\in\mathbb{R}^{(n_{\mathbf{G}}+n_{\mathbf{S}})\times (n_{\mathbf{G}}+n_{\mathbf{S}}-1)}$ and $\mathcal{H}\in\mathbb{R}^{(n_{\mathbf{G}}+n_{\mathbf{S}})\times n_{\mathbf{L}}}$ are defined as the susceptance of transmission lines matrix and the matrix related to the susceptance of transmission lines connected to the renewable energy sources and loads, respectively. Calculation of these matrices is presented in Appendix~\ref{ap:FromY2deltaW}. Observing the structure of these matrices, we can further partition them as 
\begin{equation}\label{pf600}
\mathcal{G}=\left[
\begin{array}{c|c}
\mathcal{G}_1 & \mathcal{G}_2 \\
\hline
\mathcal{G}_3 & \mathcal{G}_4
\end{array}
\right] \quad \mathcal{H}^\mathsf{T}=\bigg[\begin{array}{c|c}\mathcal{H}_1^\mathsf{T} &\mathcal{H}_2^\mathsf{T}\end{array}\bigg]^\mathsf{T},
\end{equation}
where the matrices $\mathcal{G}_1\in\mathbb{R}^{n_{\mathbf{G}}\times(n_{\mathbf{G}}-1)}$, $\mathcal{G}_2\in\mathbb{R}^{n_{\mathbf{G}} \times n_{\mathbf{S}}}$, $\mathcal{G}_3\in\mathbb{R}^{n_{\mathbf{S}}\times(n_{\mathbf{G}}-1)}$, $\mathcal{G}_4\in\mathbb{R}^{n_{\mathbf{S}} \times n_{\mathbf{S}}}$ and $\mathcal{H}_1\in\mathbb{R}^{n_{\mathbf{G}} \times n_{\mathbf{L}}}$, $\mathcal{H}_2\in\mathbb{R}^{n_{\mathbf{S}} \times n_{\mathbf{L}}}$ are with appropriate dimensions. Thereafter, revisiting the generator and the storage device models given in Sections~\ref{GM} and~\ref{SM} and considering \eqref{dc1} and~\eqref{pf600}, further calculations reveal
\begin{equation}
\begin{aligned}
\frac{\mathrm{d}}{\mathrm{d}t} \begin{bmatrix}
\delta(t)\\ \omega(t)\\ \mathsf{E}_{\mathbf{S}}(t)
\end{bmatrix}&=\begin{bmatrix}
\mathbf{0} & T & \mathbf{0} \\
-F\cdot\mathcal{G} & -\Phi & \mathbf{0} \\
\tilde{\mathcal{G}} & \mathbf{0} & \mathbf{0}
\end{bmatrix}\begin{bmatrix}
\delta(t)\\ \omega(t)\\ \mathsf{E}_{\mathbf{S}}(t)
\end{bmatrix}+\begin{bmatrix}
\mathbf{0} & \mathbf{0} \\
F & -F\cdot\mathcal{H} \\
\mathbf{0} & \mathcal{H}_2
\end{bmatrix}\begin{bmatrix}
\mathsf{P}_{\text{ref}}(t)\\
\mathsf{P}_{\mathbf{L}}(t)
\end{bmatrix} \\
&+\begin{bmatrix}
\mathbf{0} \\ \Phi \\ \mathbf{0}
\end{bmatrix}\omega_0\cdot\mathbf{1}_{(n_{\mathbf{G}}+n_{\mathbf{S}})},
\end{aligned}
\end{equation}
where $\tilde{\mathcal{G}}:=\left[\mathbf{0} \mid \mathcal{G}_4\right]$. The matrices $T\in\mathbb{R}^{(n_{\mathbf{G}}+n_{\mathbf{S}}-1)\times (n_{\mathbf{G}}+n_{\mathbf{S}})}$, $F$, $\Phi\in\mathbb{R}^{(n_{\mathbf{G}}+n_{\mathbf{S}})\times (n_{\mathbf{G}}+n_{\mathbf{S}})}$ are defined as 
\begin{align*}
T=\begin{bmatrix}
-1 & 1 & 0 & 0 & \dots \\
-1 & 0 & 1 & 0 & 0 & \dots \\
-1 & 0 & 0 & 1 & 0 & \dots \\
\vdots & 0 & 0 & 0 & \ddots & \vdots \\
-1 & 0 & 0 & 0 & 0 & 1 & \\
\end{bmatrix}, \quad F=\begin{bmatrix}
F_{\mathbf{G}} & \mathbf{0} \\
\mathbf{0} & F_{\mathbf{S}} \\
\end{bmatrix}, \quad \Phi=\begin{bmatrix}
\Phi_{\mathbf{G}} & \mathbf{0} \\
\mathbf{0} & \Phi_{\mathbf{S}} \\
\end{bmatrix},
\end{align*} 
where $F_{\mathbf{G}}$, $\Phi_{\mathbf{G}}\in\mathbb{R}^{n_{\mathbf{G}}\times n_{\mathbf{G}}}$ and $F_{\mathbf{S}},\Phi_{\mathbf{S}}\in\mathbb{R}^{n_{\mathbf{S}}\times n_{\mathbf{S}}}$ are calculated as 
\begin{align*}
F_{\mathbf{G}}&=\begin{bmatrix}
3K_1 & & \\
& \ddots & \\
& & 3K_{n_{\mathbf{G}}}
\end{bmatrix}, \quad F_{\mathbf{S}}= \begin{bmatrix}
3\frac{D_{\mathbf{S},1}}{\alpha_{\mathbf{S},1}} & & \\
& \ddots & \\
& & 3\frac{D_{\mathbf{S},n_{\mathbf{S}}}}{\alpha_{\mathbf{S},n_{\mathbf{S}}}}
\end{bmatrix}, \\
\Phi_{\mathbf{G}}&=\begin{bmatrix}
\frac{K_1}{D_{\mathbf{G},1}} & & \\
& \ddots & \\
& & \frac{K_{n_{\mathbf{G}}}}{D_{\mathbf{G},n_{\mathbf{G}}}}
\end{bmatrix}, \quad
\Phi_{\mathbf{S}}= \begin{bmatrix}
\frac{1}{\alpha_{\mathbf{S},1}} & & \\
& \ddots & \\
& & \frac{1}{\alpha_{\mathbf{S},n_{\mathbf{S}}}}
\end{bmatrix}.
\end{align*}
In this work, all the matrices stated above are computed based on the system parameters documented in Appendix~\ref{ap:Consts}.

\subsection{Bound on the total storage capacity} \label{sec:firstBound}
In this section we determine a lower bound on the total storage capacity which is required to attain the steady-state after a power transient occurs. Power transient can be considered as a disturbance, and it typically appears due to the losses of a renewable or load units. Following this, given a power network having $n_{\mathbf{L}}$ load buses, the overall power transient of the network denoted as $\mathsf{P}_{\mathrm{trans}}:\mathbb{R}_{\geq 0}\to\mathbb{R}$, can be calculated as
\begin{equation}\label{sc2}
\mathsf{P}_{\mathrm{trans}}(t):=\sum_{i=1}^{n_{\mathbf{L}}}\mathcal{P}_i(t).
\end{equation}
For the $i^{th}$ load, the step function $\mathcal{P}_i: \mathbb{R}_{\geq 0}\to\mathbb{R}$ defines a transient event. 

\begin{clm}\label{claim-1}
Given a power network, let $D_{\mathbf{G},k}$ and $D_{\mathbf{S},l}$ are the damping coefficients of the $k^{th}$ generator and the $l^{th}$ storage device, respectively, where $k\in\mathcal{N}_{\mathbf{G}}$ and $l\in\mathcal{N}_{\mathbf{S}}$. Then, to keep the steady-state frequency deviation $\Delta\omega_{ss}$ below a pre-defined value $\Delta \omega_{ss,\max}$, the minimal size of the total damping coefficients of the storage device inverters needs to be lower bounded by 
\begin{equation}\label{sc1}
\sum_{i=1}^{n_{\mathbf{S}}}\frac{1}{D_{\mathbf{S},i}}\geq\frac{3\mathsf{P}_{\mathrm{trans}}(t)}{\Delta \omega_{ss,\max}}- \sum_{i=1}^{n_{\mathbf{G}}}\frac{1}{D_{\mathbf{G},i}},
\end{equation}
where $\Delta\omega_{ss,\max}$ is the maximum allowed value of $\Delta\omega_{ss}$, while $\mathsf{P}_{\mathrm{trans}}(\cdot)$ denotes the overall power transient of the network, see~\eqref{sc2}. 
\end{clm}

A formal proof of this claim is presented in Appendix~\ref{ap:FirstBound}. Note that $\sum_{i=1}^{n_{\mathbf{S}}}\frac{1}{D_{\mathbf{S},i}}$ directly affects the total power within the storage devices, i.e., the bigger $D_{\mathbf{S},i}^{-1}$ implies more energy can be stored or used by this device. Thus, each unit's capacity size can be represent by $D_{\mathbf{S},i}^{-1}$.

\subsection{Problem formulation}\label{prfo}
The main goal of this work is to find answer to the following question: \textit{For a power system network in Section~\ref{pnm}, how to distribute $n_{\mathbf{S}}$ number of storage systems to the remaining $\left(n_{\mathbf{G}}+n_{\mathbf{L}}\right)$ number of buses such that the maximum frequency deviation will be minimized under transient events?} This problem can be formulated as an optimization problem stated below
\begin{equation} \label{mainopt}
\begin{aligned}
& \underset{\mathcal{D}_{\mathbf{S}}}{\text{minimize}}
& & |\omega_0-\omega_{\mathrm{nadir}}(\mathcal{D}_{\mathbf{S}})| \\
& \text{subject to}
& & \sum_{i=1}^{n_{\mathbf{S}}}D_{\mathbf{S},i}^{-1}=D_{\mathbf{S},\mathrm{total}}^{-1}, \\
&&& D_{\mathbf{S},i}^{-1}=\frac{1}{n_{\mathbf{S}}}D_{\mathbf{S},\mathrm{total}}^{-1}\quad \forall i\in\mathcal{N}_{\mathbf{S}},\\
&&& D_{\mathbf{S},\mathrm{total}}^{-1}=\frac{3\mathsf{P}_{\text{trans}}(t)}{\Delta \omega_{ss,\max}}- \sum_{i=1}^{n_{\mathbf{G}}}\frac{1}{D_{\mathbf{G},i}},\\
&&& D_{\mathbf{S},\mathrm{total}}^{-1}\in\mathbb{Z}_{+},\\
&&& n_{\mathbf{S}}\in\mathbb{Z}_{+},\quad n_{\mathbf{S}}\leq\left(n_{\mathbf{G}}+n_{\mathbf{L}}\right). 
\end{aligned}
\end{equation}
Here, the set $\mathcal{D}_{\mathbf{S}}$ is defined as all combinations of $\mathcal{D}_{\mathbf{S}}:=\left\lbrace D_{\mathbf{S},i}~|~i\in\mathcal{N}_{\mathbf{S}}\right\rbrace$ and $D_{\mathbf{S},\mathrm{total}}$ denotes the size of the total damping coefficients of the storage devices. The term $\omega_{\mathrm{nadir}}$ is defined as the maximum change of generators' frequencies on the time domain, and it is represented as
\begin{equation}\label{eq:FreqNadir}
\omega_{\mathrm{nadir}}= \begin{cases}
\max\limits_{t\geq 0}\max\limits_{i\in\mathcal{N}_{\mathbf{G}}}|\omega_{\mathbf{G},i}(t)|,\quad \text{if}~\omega_{\mathbf{G}}(t)>\omega_{0}\cdot\mathbf{1}_{n_{\mathbf{G}}}, \\
\min\limits_{t\geq 0}\min\limits_{i\in\mathcal{N}_{\mathbf{G}}}|\omega_{\mathbf{G},i}(t)|,\quad \text{if}~\omega_{\mathbf{G}}(t)<\omega_{0}\cdot\mathbf{1}_{n_{\mathbf{G}}}.
\end{cases} 
\end{equation}
From the optimization problem~\eqref{mainopt} it can be easily noticed that the capacity of all the storage units are equal and the total size is defined based on the maximum allowed frequency change at steady-state as shown in \eqref{sc1}. This problem can be considered as a combinatorial problem which consists of a combination of $|\mathcal{D}_{\mathbf{S}}| =\binom{n_{\mathbf{G}}+n_{\mathbf{L}}+n_{\mathbf{S}}-1}{n_{\mathbf{S}}}$ solutions, since each location (bus) can have more than single storage unit.

\section{Storage units distribution for frequency stability: Numeric approaches} \label{sec:numericAppr}
In this section we aim to solve problem~\eqref{mainopt} by exploiting combinatorial optimization methods. To this end, Section~\ref{bfs} describes a solution to this problem employing `Brute-force search' method, while a solution based on the adaptation of `Cross-entropy' method is presented in Section~\ref{cem}. 

\subsection{Solution based on the Brute-force search} \label{bfs}
Since the problem~\eqref{mainopt} is discrete and all the variables are integers, it can be addressed employing the `brute-force algorithm'. This algorithm typically searches for the optimal solution out of all optional solutions subject to the constrains given in \eqref{mainopt}. It requires the total number of storage units $n_{\mathbf{S}}$ and the total capacity of the storage devices $D_{\mathbf{S},\mathrm{total}}^{-1}$ as inputs, and it provides the best energy storage distribution over the network. First, all the optional combinations of the storage unit distributions are created. Then, each combination is tested by different transient scenarios. Once all the combinations are tested, the best storage distribution policy is selected. The key steps of this algorithm are presented in Algorithm~\ref{ProposedAlgorithmProblem1}. Although this algorithm provides best storage distribution over the network, it is computationally complex. For instance, given a power network, if the number of buses in the transmission network and the number of energy storage systems are increased then computational complexity of the algorithm increases accordingly.

\begin{algorithm}[H]\label{ProposedAlgorithmProblem1}
\caption{Storage distribution policy using Brute-force search}
\begin{algorithmic}[1]
\State \textbf{Data:}~$n_{\mathbf{S}}$, $D_{\mathbf{S},\mathrm{total}}^{-1}$,
\State \textbf{Result:}~\textit{Best distribution},
\State Select \textit{Lowest cost} = $\infty$,
\State Create all combinations for storage distribution, i.e., $\mathcal{D}_{\mathbf{S}}$,
\State Set $d=1$,
\While {($d \leq |\mathcal{D}_{\mathbf{S}}|$)}
\State Run all optional transient events,
\State Calculate $\omega_{\mathrm{nadir}}(d)$ for each event,
\State Calculate \textit{cost function}($d$) \eqref{mainopt},
\If {$( \textit{cost~function}(d) \leq \textit{Lowest~Cost} )$}
\State Set \textit{Lowest~cost} := \textit{cost~function}($d$),
\State Declare \textit{Best~distribution}=$d$,
\EndIf
\State Update $d=d+1$,
\EndWhile
\State Return:~\textit{Best distribution}
\end{algorithmic}
\end{algorithm}

\subsection{Solution based on Cross-Entropy method} \label{cem}
The combinatorial optimization problem in~\eqref{mainopt} can also be formulated as
\begin{equation}\label{c_mainopt}
C(\hat{x})=\hat{\gamma}= \underset{x\in\mathcal{D}_{\mathbf{S}}}{\text{minimize}}~|\omega_0-\omega_{\mathrm{nadir}}(x)|,
\end{equation}
which is subject to the same constraints given in~\eqref{mainopt}. In~\eqref{c_mainopt}, $C:\mathcal{D}_{\mathbf{S}}\to\mathbb{R}_{\geq 0}$ is the performance function, $\mathcal{D}_{\mathbf{S}}$ is a discrete set and $\hat{x}$ is the optimal solution. We observe in Section~\ref{bfs} that if the number of elements in $\mathcal{D}_{\mathbf{S}}$ are increased, the computational effort needs to solve~\eqref{c_mainopt} also increases significantly. To tackle this problem, in the sequel, we employ adaptation to the `Cross Entropy' (CE) method to solve~\eqref{c_mainopt}. The CE method initially proposed to efficiently estimate rare-event probabilities \cite{CE1}, and later extended to solve combinatorial optimization problems \cite{CE2}. This method defines a precise mathematical framework for evaluating fast update and learning rules. It is used successfully in several fields, including power systems and smart grids \cite{CePower1,CePower2}. 

We consider $x\in \mathcal{D}_{\mathbf{S}}$ as a random variable with a probability mass function (PMF) $f:\mathcal{D}_{\mathbf{S}}\to\mathcal{B}$\footnote{The set $\mathcal{B}$ collectively represents all the generator and load buses i.e. $\mathcal{B}:=\left\lbrace \mathcal{B}_1,\mathcal{B}_2,\ldots,\mathcal{B}_n\right\rbrace$ where $n=n_{\mathbf{G}}+n_{\mathbf{L}}$. The number of storage device placed at the $i^{th}$ bus is denoted as $\mathcal{B}_i=0$ if it has no storage, $\mathcal{B}_i=b_i$ otherwise, where $b_i\in\left[1,n_{\mathbf{S}}\right]$. Note that $\sum_{i=1}^nb_i=n_{\mathbf{S}}$.}. Furthermore, for a real number $\gamma$, we define $G:\mathcal{D}_{\mathbf{S}} \times \mathbb{R}\to\left\lbrace 0,1\right\rbrace$ as
\begin{equation}\label{eq:Gx_gama}
G(x,\gamma):= \begin{cases}
1, \quad \text{if} & C(x)\leq\gamma, \\
0, \quad \text{if} & C(x)>\gamma.
\end{cases} 
\end{equation}
The probability for which $\left( C(x)\leq\gamma\right)$ is considered as 
\begin{equation}
\mathsf{g}(\gamma)=\mathbb{E}\left[G(x,\gamma)\right]. \label{eq:meanG}
\end{equation}
Since $\mathsf{g}(\cdot)$ is unknown, and for $\gamma=\hat{\gamma}$, the probability \eqref{eq:meanG} is very small, it is considered as a \textit{rare-event}. Therefore, an estimate of $\mathsf{g}(\cdot)$ denoted as $\mathsf{\hat{g}}(\cdot)$, can be calculated by the next average as
\begin{equation}
\mathsf{\hat{g}}(\gamma)=\sum_{k=1}^MG(x_k,\gamma)\frac{f(x_k)}{h(x_k)},
\end{equation} 
where $M\in\mathbb{Z}_{+}$ and $h:\mathcal{D}_{\mathbf{S}}\to\mathcal{B}_n$ is a known PMF of $\mathcal{D}_{\mathbf{S}}$. This technique is termed as the \textit{importance sampling technique} \cite{ImportanceSampl} and the values $x_k$ are considered as random samples of $\mathcal{D}_{\mathbf{S}}$. An optimal (zero variance) method to estimate $\hat{\mathsf{g}}(\cdot)$ is to use the ideal importance sampling PMF, which is given by
\begin{equation}
\hat{h}(x)=\frac{G(x_k,\gamma)f(x)}{\mathsf{g}(\gamma)}. \label{eq:hIdeal}
\end{equation}
Here, $\hat{h}(\cdot)$ is considered to be optimal if most of the probability mass is assigned close to $\hat{x}$. The difficulty here is that $\hat{h}(\cdot)$ depends on the unknown parameter $\mathsf{g}(\cdot)$. To overcome this, the CE method searches in $\mathsf{H}$ the element $h(\cdot)$ which distance from the ideal sampling distribution is minimal, where $\mathsf{H}$ is a given set of PMFs.

The CE method aims to estimate the optimal PMF by adaptively selecting members $h(\cdot)$ of $\mathsf{H}$ that are closest to $\hat{h}(\cdot)$ in the sense of the Kullback-Leibler divergence. This measure is also termed the cross-entropy between $\hat{h}(\cdot)$ and $h(\cdot)$. Thus, the problem then reduces to
\begin{equation}\label{eq:KLargmin}
\argmin_{h \in \mathsf{H}}\mathrm{KL}(\hat{h},h)=
\argmin_{h \in \mathsf{H}, x\in \mathcal{D}_{\mathbf{S}}}
\mathbb{E}\left[\log\frac{\hat{h}(x)}{h(x)}\right].
\end{equation}

Note that since the value of $C(\hat{x})$ is unknown, random samples $\mathcal{X}_1 \subset \mathcal{D}_{\mathbf{S}} $ can be selected as inputs to the algorithm, and then $\gamma_1= \min_{x \in \mathcal{X}_1 }{C(x)}$ is calculated. Afterwards, problem \eqref{c_mainopt} is solved using \eqref{eq:KLargmin}, where the set $\mathsf{H}$ is an input to the algorithm and $h_1(\cdot)$ is used in the first iteration. Following this, $h_{\mathrm{iter}}(\cdot)$ are computed iteratively. Assuming that the number of random samples per iteration is large enough, and $h_{\mathrm{iter}}(\cdot)$ is not too far from the ideal sampling distribution, these PMFs become more likely to generate samples that have elements corresponding to low-values of $C(\cdot)$, when the number of iteration increases.

The iterative procedure for energy storage distribution can be divided into two phases per iteration:
\begin{enumerate}
\item Each solution in the algorithm is described by the set $\mathcal{B}_n$ under the constraint that the total size of the vector is $n_{\mathbf{S}}$. The value $b_i\in\left[1,n_{\mathbf{S}}\right]$ is generated according to probability metric $Q$ which has a Bernoulli distribution. Each value in the metric $Q$ represents the probability to locate a storage unit in the bus of that index and the initialize probability is uniform across the metric such that $Q[\mathrm{iter}=1] = [q_1,q_2,\ldots,q_n] = [{\frac{1}{n}, \frac{1}{n},\ldots, \frac{1}{n}}]$. This randomization is done $n_{\mathbf{S}}$ times. 

For example: $n_{\mathbf{S}} =3$ and $n= 5$, each solution will include 3 raffles of a number between 1--5 with the probability of $\frac{1}{5}$ in the first iteration. For a specific solution, if the raffles are $[4,2,4]$ then the solution is $\{0,1,0,2,0\}$, i.e., single storage unit in bus 2 and two storage units in bus 4.
\item At each iteration, $ |X_{s_{CE}}|$ samples are randomized from $| \mathcal{D}_{\mathbf{S}}|$ using the $Q[\mathrm{iter}]$ metric. At the end of each iteration the best $\epsilon$ solutions are choosen in order to update the probabilities in $Q[\mathrm{iter}+1]$ metric employing the following relation
\begin{equation}\label{eq:qUpdate}
q_i[\mathrm{iter}+1] = \frac{\beta\mathcal{O}_i[\mathrm{iter}]}{\epsilon|X_{s_{CE}}| } + (1-\beta)q_i[\mathrm{iter}+1] ,
\end{equation}
where $i\in (1,n)$ represent the bus index, $O_i[\mathrm{iter}]$ is the number of times a storage unit was placed at bus $i$ in the elite group at iteration $\mathrm{iter}$ and $\beta$ is the smoothing factor.
\end{enumerate} 

The process is summarized as follows in Algorithm~\ref{ProposedAlgorithmProblem2}.\\
\begin{algorithm}[H]\label{ProposedAlgorithmProblem2}
\caption{Storage distribution policy using the CE method}
\begin{algorithmic}[1]
\State \textbf{Input data:}~$n_{\mathbf{S}}$, $D_{\mathbf{S},\mathrm{total}}^{-1}$, $N_{\mathrm{iter}}$, $|X_{s_{CE}}|$, $\epsilon$, $\beta$,
\State \textbf{Output data:}~$Q[N_{\mathrm{iter}}]$, \textit{Best distribution},
\State Initialize probability metric $Q[\mathrm{iter}=1]=[{\frac{1}{n},\frac{1}{n},\ldots,\frac{1}{n}}]$,
\State Set \textit{Lowest Cost} := $\infty$,
\While {($\mathrm{iter}\leq N_{\mathrm{iter}}$)}
\State Randomize $|X_{s_{CE}}|$ samples from $|\mathcal{D}_{\mathbf{S}}|$ according to the probability metric $Q[\mathrm{iter}]$,
\For {each randomized samples $d$}
\State Run on all optional transient events,
\State Calculate $\omega_{\mathrm{nadir}}(d)$ for each event,
\State Calculate \textit{cost function}($d$) \eqref{mainopt},
\If {$(\textit{cost~function}(d) \leq \textit{Lowest~Cost})$}
\State Set \textit{Lowest Cost} := \textit{cost function}($d$),
\State Select \textit{Best Distribution} :=$d$
\EndIf
\EndFor
\State Select the $\epsilon$ $|X_{s_{CE}}|$ best samples based on the cost function. These are named the ``elite group'',
\State For each sample in the elite group, collect statistics from $O_b[\mathrm{iter}]$,
\State Update $Q[\mathrm{iter}+1]$ based on \eqref{eq:qUpdate},
\State Set $\mathrm{iter}:=\mathrm{iter}+1$,
\EndWhile
\State Return:~$Q[N_{\mathrm{iter}}]$, \textit{Best distribution}.
\end{algorithmic}
\end{algorithm}

The problem formulation can be considered as combinatorial problem which consists of $N_{\mathrm{iter}}|X_{s_{CE}}|$ solutions, where $N_{\mathrm{iter}}$ is the number of iterations and $|X_{s_{CE}}|$ is the number of random solutions per iteration (constant).
In this work the next definition for comparing computational complexity between the approaches is suggested as
\begin{equation}\label{ratio}
\text{complexity ratio} = \frac{|\text{ solutions(approach 1)}|}{| \text{ solution(approach 2)}|} = \frac{\binom{n_{\mathbf{G}}+n_{\mathbf{L}}+n_{\mathbf{S}}-1}{n_{\mathbf{S}}}}{N_{\mathrm{iter}}|X_{s_{CE}}|}.
\end{equation}

\section{The future Israeli grid: Case study } \label{sec:caseStudy}

In this section we perform a series of numerical experiments on the Israeli electricity grid to validate the algorithmic strategies shown in Algorithms~\ref{ProposedAlgorithmProblem1} and~\ref{ProposedAlgorithmProblem2}.

\subsection{Future Israeli grid: Simplified model}

\begin{figure}[htbp]
\centering
\includegraphics[height=0.5\textheight]{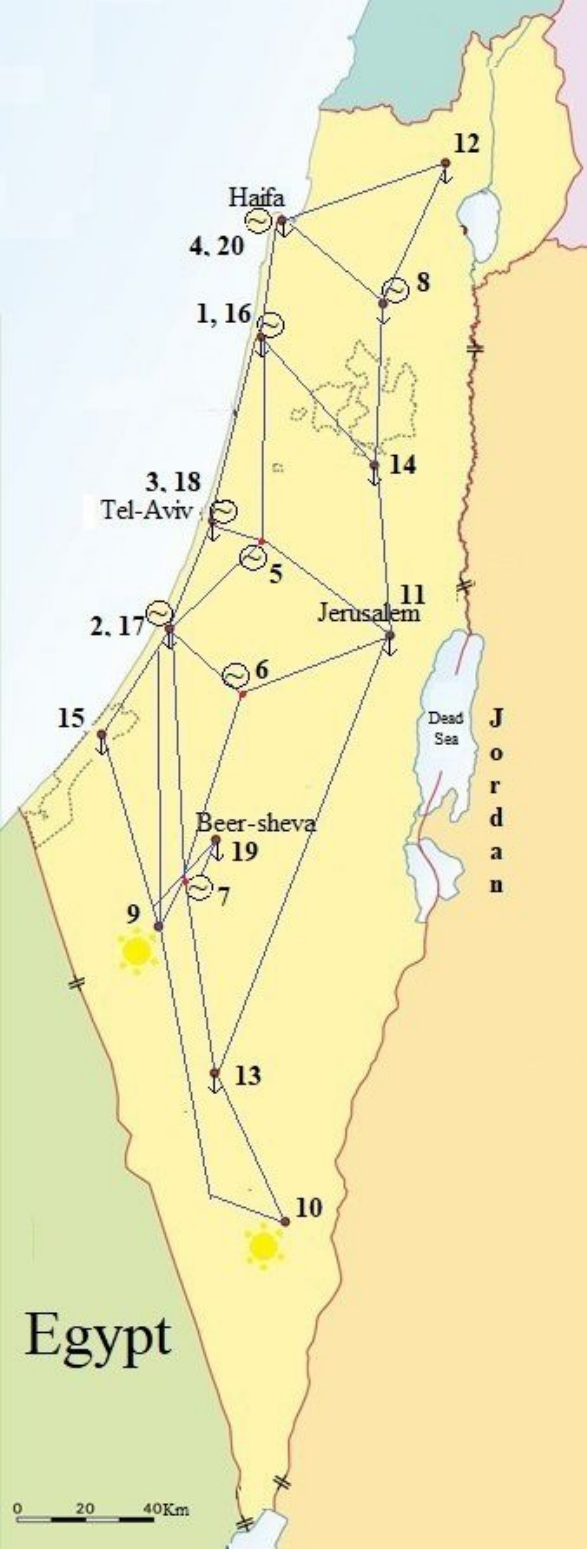}
\caption{ Simplified version of the future Israeli electricity grid in $2025$.}
\label{fig:IsraelLoadGenTran}
\end{figure}

We consider the future Israeli electricity grid~\cite{Israel2025} shown in Fig.~\ref{fig:IsraelLoadGenTran}. In this grid, an amount of $18\%$ of the generated electricity is came from the centralized solar fields, which are placed at the south of Israel. Reliability of the transmission system is a prime consideration in Israel for two main reasons: (i) Israel does not have interconnections with other electrical networks and (ii) since the state of Israel is small, the grid is more sensitive to disturbances~\cite{Navon2020}. Thus, security is the main design criteria. Security addresses the ability of the system to survive failures without losing the ability to supply electricity to consumers. Since we are focusing on the power system stability using storage units, we examine different failure scenarios related to renewable power generations loss in the subsequent simulations. We consider the grid in Fig.~\ref{fig:IsraelLoadGenTran} with 20 buses which include 8 generator buses and 12 load buses. Among the load buses, two solar fields considered as negative loads, are connected to bus 9 and 10. The total consumption and production of this grid is 10.016 GWs. Further, we assume $V_{\mathrm{base}} = 400$ KV and $P_{\mathrm{base}} = 100$ MVA.

\subsection{Model validation: Placement of single storage unit}
\begin{figure}[htbp]
\centering
\includegraphics[width=0.7\textwidth]{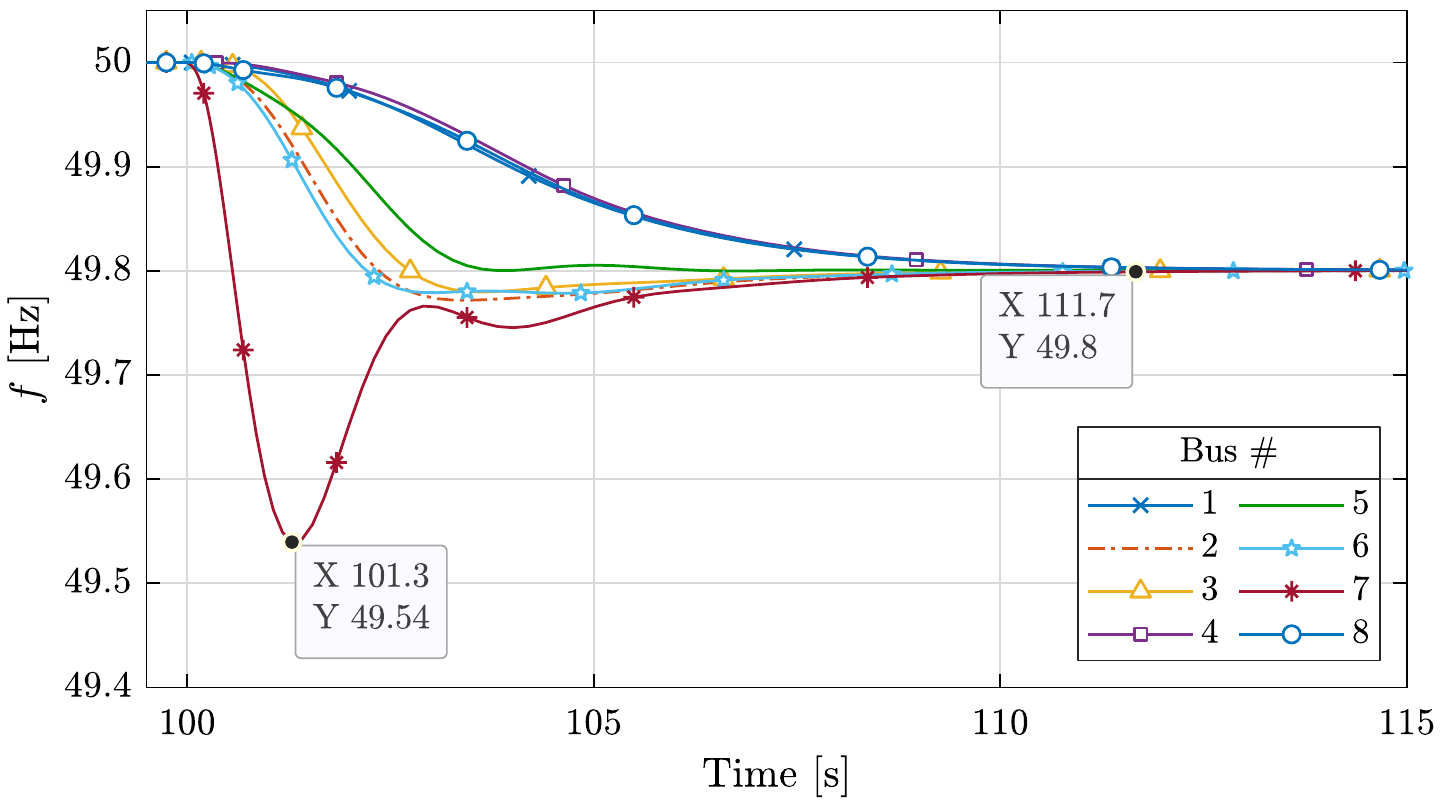}
\caption{Frequency evolution of synchronous generators considering the scenario when the $\mathsf{P}_{\mathrm{trans}}$ occurs at bus 10 and the storage unit is also connected with the same bus.}
\label{fig:resultlema1}
\end{figure}

First, we validate Claim~\ref{claim-1}. We consider the grid in Fig.~\ref{fig:IsraelLoadGenTran}, and assume a storage unit is connected to the bus 10. Furthermore, we consider a power transient which emulates a renewable power generation loss of 1100 MW, also occurs at the same bus. From the Israeli model data-sheet \cite{Israel2025}, we collect data of all the $D_{\mathbf{G},i}$ where $i\in\mathcal{N}_{\mathbf{S}}$, and consider $\mathsf{P}_{\mathrm{trans}}=1100$ MW and $\omega_{ss,\max}=2\pi49.8$ $\frac{\text{rad}}{\text{sec}}$. Then, based on these data and employing \eqref{sc1}, we obtain $\frac{1}{D_{\mathbf{S},10}} = 240$ MWs. The frequency evolution of all the generators are depicted in Fig.~\ref{fig:resultlema1}, and it is observed that the steady-state frequencies (denoted as $f_{ss}$) of each generator is 49.8~Hz, as expected.

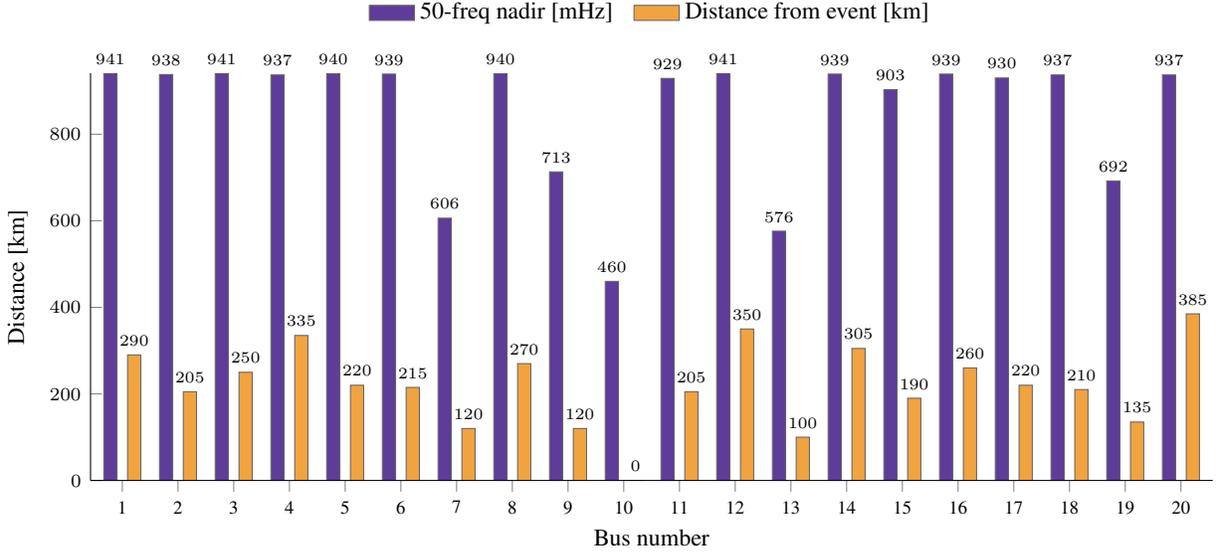
\begin{figure}[htbp]
\centering
\begin{tikzpicture}[font=\scriptsize]
\begin{axis}[
height=7cm,
width=16.5cm,
enlarge y limits={upper, value=0},
enlarge x limits = 0.03,
ybar=4,
bar width = 5pt,
symbolic x coords={1,2,3,4,5,6,7,8,9,10,11,12,13,14,15,16,17,18,19,20},
x tick label style={align=center},
xtick=data,
nodes near coords,
every node near coord/.append style={font=\tiny,anchor=south},
xlabel={Bus number},
x label style={font=\footnotesize},
ylabel={Distance [km]},
y label style={at={(axis description cs:0.035,.5)},anchor=south,font=\footnotesize},
axis lines*=left,
cycle list={
{fill=COL1,draw=black!60},
{fill=COL2,draw=black!60},
},
axis on top,
legend style={draw=none,at={(0.5,1.2)},font=\footnotesize,anchor=north,/tikz/every even column/.append style={column sep=0.2cm}},
legend cell align={left},
legend columns=-1,
area legend
]
\addplot coordinates {(1,941) (2,938) (3,941) (4,937) (5,940) (6,939) (7,606) (8,940) (9,713) (10,460) (11,929) (12,941) (13,576) (14,939) (15,903) (16,939) (17,930) (18,937) (19,692) (20,937)};
\addplot coordinates {(1,290) (2,205) (3,250) (4,335) (5,220) (6,215) (7,120) (8,270) (9,120) (10,0) (11,205) (12,350) (13,100) (14,305) (15,190) (16,260) (17,220) (18,210) (19,135) (20,385)};
\addlegendentry{50-freq nadir [mHz]}
\addlegendentry{Distance from event [km]}
\end{axis}
\end{tikzpicture}
\caption{$|f_0-f_{\mathrm{nadir}}|$ and the distance of the storage units from the bus 10.}
\label{fig:plotHzKmBusN}
\end{figure}

Table~\ref{table:storageDiffrentLocations} documents frequency nadir (denoted as $f_{\mathrm{nadir}}$) and $f_{ss}$, considering the scenarios where the $\mathsf{P}_{\mathrm{trans}}$ occurs at bus 10, and the storage is placed at different locations. From Table~\ref{table:storageDiffrentLocations}, we found that $f_{ss}$ is same for all the locations, while $f_{\mathrm{nadir}}$ is location dependent. In particular, the maximum of $f_{\mathrm{nadir}}$ is achieved when the storage is placed at bus 10 where the transient occurs. In Fig.~\ref{fig:plotHzKmBusN} we compute the cost function in~\eqref{mainopt} and calculate the distance of the storage units from the bus 10 where the transient occurs. It can be easily verified that when the storage unit is placed near the bus 10, $f_{\mathrm{nadir}}$ is increased significantly which cause the objective function in \eqref{mainopt} to decrease.

\begin{table}[htbp]
\centering
\caption{Values of $f_{\mathrm{nadir}}$ and $f_{ss}$ based on the storage device locations considering power transient occurs at bus 10.}
\label{table:storageDiffrentLocations}
\renewcommand{\arraystretch}{1.1}
\begin{tabular}{l | c c} 
\toprule
Storage connected to & $f_{\mathrm{nadir}}$ [in Hz] & $f_{ss}$ [in Hz] \\
\midrule
No storage & $49.0544$ & $49.782$\\ 
Bus $1$ & $49.0594$ & $49.8$ \\
Bus $2$ & $49.0615$ & $49.8$ \\
Bus $3$ & $49.0595$ & $49.8$ \\
Bus $4$ & $49.0631$ & $49.8$ \\
Bus $5$ & $49.0598$ & $49.8$ \\
Bus $6$ & $49.0613$ & $49.8$ \\
Bus $7$ & $49.3936$ & $49.8$ \\
Bus $8$ & $49.0601$ & $49.8$ \\
Bus $9$ & $49.2865$ & $49.8$ \\
Bus $10$ & $\mathbf{49.5396}$ & $49.8$ \\
Bus $11$ & $49.0714$ & $49.8$ \\
Bus $12$ & $49.0594$ & $49.8$ \\
Bus $13$ & $49.4235$ & $49.8$ \\
Bus $14$ & $49.0608$ & $49.8$ \\
Bus $15$ & $49.0973$ & $49.8$ \\
Bus $16$ & $49.0609$ & $49.8$ \\
Bus $17$ & $49.0698$ & $49.8$ \\
Bus $18$ & $49.0627$ & $49.8$ \\
Bus $19$ & $49.3076$ & $49.8$ \\
Bus $20$ & $49.0633$ & $49.8$ \\		
\bottomrule
\end{tabular}
\end{table}

\subsection{Placement of storage units using Brute-force search} \label{subsec:BF5ESS}
\begin{figure}[htbp]
\centering
\includegraphics[width=0.7\textwidth]{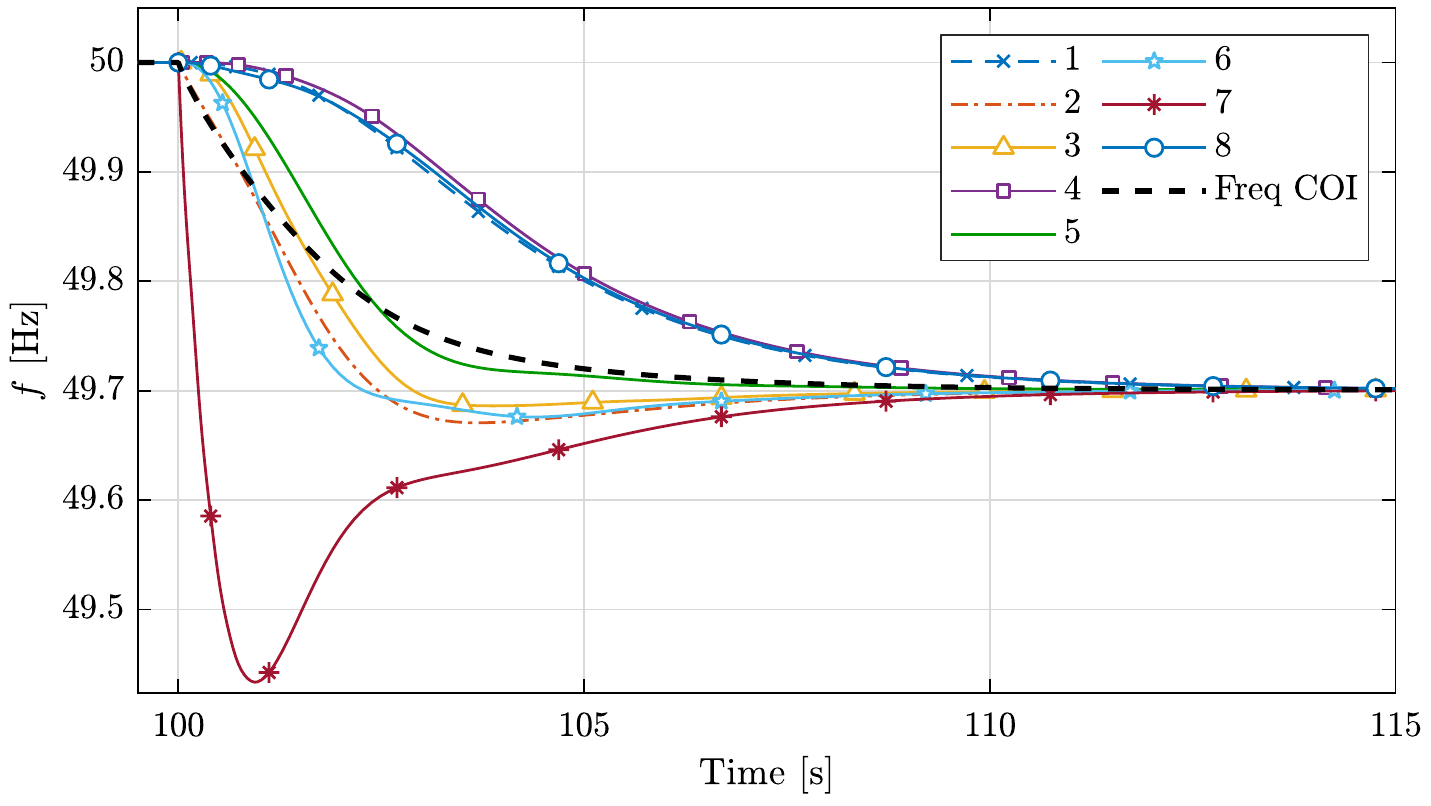}
\caption{Best distribution: $\left\lbrace \mathbf{Bus},\mathsf{Storage}\right\rbrace=\{\mathbf{7},\mathsf{2}\}, \{\mathbf{10},\mathsf{3}\}$.}
\label{fig:bestCase}
\end{figure}

\begin{figure}[htbp]
\centering
\includegraphics[width=0.7\textwidth]{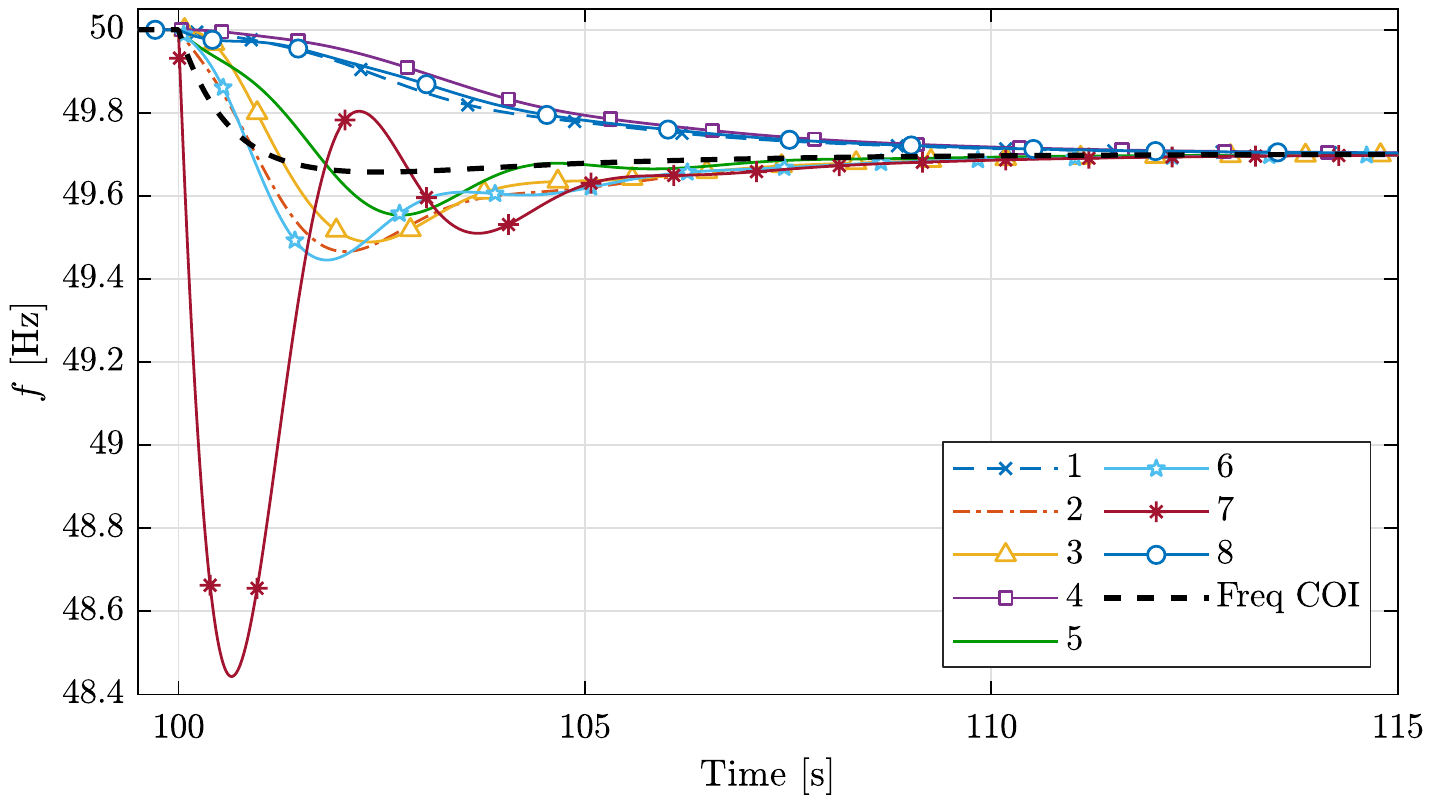}
\caption{Worst distribution: $\left\lbrace \mathbf{Bus},\mathsf{Storage}\right\rbrace=\{\mathbf{1},\mathsf{1}\}, \{\mathbf{4},\mathsf{4}\}$.}
\label{fig:worstCase}
\end{figure}

First, we aim to place $n_{\mathbf{S}}=5$ storage units in this grid to minimize $f_{\mathrm{nadir}}$ employing the Brute-force search method . The Israel Electric Corporation allows up to 0.3~Hz deviation from the nominal frequency, which implies $\omega_{ss,\max}=2\pi49.7$ rad/s. We assume the power transient emulates a renewable power generations loss of 1.8 GW in both buses 9 and 10, ($\mathsf{P}_{\mathrm{trans}}=1.8$~GW). Let $\sum_{i=1}^{n_{\mathbf{G}}}D_{\mathbf{G},i}^{-1}=2.3$ GWs, then employing Claim~\ref{claim-1} we obtain $D_{\mathbf{S},\mathrm{total}}^{-1}=480$ MWs. Considering the constraints in~\eqref{mainopt}, we set $D_{\mathbf{S},i}^{-1}=96$ MWs for each $i\in\left\lbrace 1,2,\ldots,5\right\rbrace$ and found that the problem consists of $|\mathcal{D}_\mathbf{S}|=\binom{5+20-1}{5} = 42504$ solutions. The experiments are performed in the Matlab/Simulink environment on the Intel i7 1.9 GHz laptop with 16 of GB RAM, and the run-time is 10.26 hours. In Table~\ref{table:IsraelStorageDist}, six best distributions are presented, in which the best distribution is two storage units at bus 7 and three units at bus 10, i.e., $\{\mathbf{7},\mathsf{2}\}$, $\{\mathbf{10},\mathsf{3}\}$\footnote{Throughout this simulation, given two positive integers $\mathbf{c},\mathsf{d}\in\mathbb{Z}_{+}$, the symbol $\left\lbrace \mathbf{c},\mathsf{d}\right\rbrace$ denotes that $\mathsf{d}$ number of storage units are placed at bus $\mathbf{c}$.}. The corresponding $f_{\mathrm{nadir}}=49.4336$ Hz, and it is depicted in Fig.~\ref{fig:bestCase}. Rest of the five best distributions show that at least one storage is placed at bus 7 and at least two storage units are placed at bus 10. These distributions describe that the best options to locate the storage systems are buses near or at the buses where the transient occurs. From Table~\ref{table:IsraelStorageDist} it can also be concluded that the worst distributions of the storage units are located at the center and the north of Israel, which is far away from the disturbances in the south. One of these solutions is shown in Fig.~\ref{fig:worstCase} where one storage unit is placed at bus 1 and the other four units are placed at bus 4 ($\{\mathbf{1},\mathsf{1}\},\{\mathbf{4},\mathsf{4}\}$), see Table~\ref{table:IsraelStorageDist}. For this distribution, $f_{\mathrm{nadir}}= 48.443$ Hz. Therefore, we notice that the difference of $f_{\mathrm{nadir}}$ between the best and worst distributions is almost 1~Hz.

In Table \ref{table:IsraelStorageDist} and Figs.~\ref{fig:bestCase} and \ref{fig:worstCase} we compute $f_{\mathrm{nadir}}$ and $f_{\mathrm{coi},\min}$. It needs to be remarked that the $f_{\mathrm{coi}}$ which is used in~\cite{SizeLocation_2_Mexican2018,SizeLocation_freqConstrains_2019,SizeLocation_1_2016}, is less accurate for inertial response, since during the transient event the generators' frequencies are not equal. The results in Table~\ref{table:IsraelStorageDist} are aligned with the above claim, the variance of $f_{\mathrm{coi},\min}$ for all optional distributions is between 49.7007 and 49.6525 Hz, less than 0.05 Hz difference between the maximum and minimum frequency deviation. Furthermore, we observe that there are 18977 optional distributions for which the $f_{\mathrm{coi},\min}$ reaches to its minimum deviation at 49.7007 Hz.

The above case study provides a new insight regarding the locations of storage units in case of transient event. While other works suggest that the locations should be in the areas of low inertia \cite{Location_FlywheelChile2017} and that the locations of the disturbance and storage effect the resilience of the grid more than the total inertia \cite{Location_ETH_PlaceVirtualInertia2017}, our simulation results indicate that the storage can be located also in sites with high inertia around the area of the disturbances in accordance to the network topology. As shown in Table~\ref{table:IsraelStorageDist} all the best distributions contain at least one storage at bus 7 which contains synchronous generator with high inertia that generate almost $9\%$ of the entire power in the grid. Since it considers as a central bus of the grid which connected to other five buses, in case of a disturbance around it more inertia is required and thus energy storage should be located in this bus as well.

\begin{table}[htbp]
\centering
\renewcommand{\arraystretch}{1.1}
\caption{Value of $f_{\mathrm{nadir}}$ under transient at all renewable resources simultaneously when 5 storage devices (each with $D_{\mathbf{S},i}^{-1}=96$ MWs) are distributed for all optional combinations.}
\label{table:IsraelStorageDist}
\begin{tabular}{c | l c c c} 
\toprule
\# & Distribution & $f_{\mathrm{nadir}}$ [in Hz] & $f_{\mathrm{coi},\min}$ [in Hz] & $\left\{\mathbf{Bus},\mathsf{Storage}\right\}$ \\
\midrule
$1$ & Best distribution & $49.4336$ & $49.7007$ & $\{\mathbf{7},\mathsf{2}\}$, $\{\mathbf{10},\mathsf{3}\}$ \\ 
$2$ & $2^{nd}$ best distribution & $49.4162$ & $49.7007$ & $\{\mathbf{7},\mathsf{1}\}$, $\{\mathbf{10},\mathsf{3}\}$, $\{\mathbf{13},\mathsf{1}\}$ \\ 
$3$ & $3^{rd}$ best distribution & $49.4159$ & $49.7007$ & $\{\mathbf{7},\mathsf{1}\}$, $\{\mathbf{10},\mathsf{3}\}$, $\{\mathbf{19},\mathsf{1}\}$ \\ 
$4$ & $4^{th}$ best distribution & $49.4126$ & $49.7007$& $\{\mathbf{7},\mathsf{3}\}$, $\{\mathbf{10},\mathsf{2}\}$ \\ 
$5$ & $5^{th}$ best distribution & $49.4101$ & $49.7007$ & $\{\mathbf{7},\mathsf{2}\}$, $\{\mathbf{10},\mathsf{2}\}$, $\{\mathbf{13},\mathsf{1}\}$ \\ 
$6$ & $6^{th}$ best distribution & $49.4058$ & $49.7007$ & $\{\mathbf{7},\mathsf{1}\}$, $\{\mathbf{10},\mathsf{4}\}$ \\ 
$\vdots $ &	$\vdots $ & $\vdots $ & $\vdots$ & $\vdots$\\ 
$38195$ & Worst $f_{\mathrm{coi},\min}$ & $48.4469$ & $49.6525$ & $\{\mathbf{12},\mathsf{5}\}$ \\ 
$\vdots $ &	$\vdots $ & $\vdots $ & $\vdots$ & $\vdots $\\ 
$42401$ & Worst distribution & $48.443$ & $49.6815$ & $\{\mathbf{1},\mathsf{2}\}$, $\{\mathbf{3},\mathsf{2}\}$, $\{\mathbf{4},\mathsf{1}\}$ \\ 
$\vdots $ &	$\vdots $ & $\vdots $ & $\vdots$ & $\vdots $\\ 
$42504$ & Worst distribution & $48.443$ & $49.6576$ & $\{\mathbf{1},\mathsf{1}\}$, $\{\mathbf{4},\mathsf{4}\}$ \\ 
\bottomrule
\end{tabular}
\end{table}

\subsection{Placement of storage units based on CE methods}
From Section~\ref{subsec:BF5ESS} we observe that the Brute-force algorithm takes a lot of time to converge to the optimal solution for a small number of storage units (for $n_{\mathbf{S}}=5$, the run-time is 10.26 hours). Therefore, to avoid this shortcoming, here we employ the CE method to solve~\eqref{mainopt}. We consider four scenarios where we place different number of storage units (for instance $n_{\mathbf{S}}=5,8$ and $10$) to the Israeli grid shown in Fig.~\ref{fig:IsraelLoadGenTran}. The values of $\omega_{ss,\max}$, $\mathsf{P}_{\mathrm{trans}}(t)$ and $\sum_{i=1}^{n_{\mathbf{G}}}D_{\mathbf{G},i}^{-1}$ are same as considered in Section~\ref{subsec:BF5ESS} and $D_{\mathbf{S},\mathrm{i}}^{-1}$ for each storage unit is calculated accordingly. The experimental data for all the scenarios are documented in Table~\ref{table:IsraelStorageDistAlg2Scenrio}. Note that, the third and fourth scenario consist of the same number of storage units but with different simulation parameters. 

\begin{table}[htbp]
\centering
\renewcommand{\arraystretch}{1.1}
\caption{Experimental test data for each scenarios.}
\label{table:IsraelStorageDistAlg2Scenrio}
\begin{tabular}{c | c c c c c c c} 
\toprule
Scenario & $n_{\mathbf{S}}$ &$D_{\mathbf{S},i}^{-1}$& $\beta$ & $\epsilon$ & $N_{\mathrm{iter}}$ & $|X_{s_{CE}}|$ & Complexity-ratio \eqref{ratio}\\
\midrule
$1$ & $5$ & $96$ & $0.03$ & $0.125$ & $20$ & $150$ & $14.17$ \\ 
$2$ & $8$ & $60$ & $0.03$ & $0.125$ & $30$ & $250$ & $296$\\
$3$ & $10$ & $48$ & $0.03$ & $0.125$ & $30$ & $250$ & $2670.67$\\
$4$ & $10$ & $48$ & $0.03$ & $0.125$ & $35$ & $300$ & $1907.6$ \\
\bottomrule
\end{tabular}
\end{table}

For each scenarios, Table~\ref{table:IsraelStorageDistAlg2} documents the best distribution and its corresponding $f_{\mathrm{nadir}}$. We observe that for the first scenario, the best distribution and the $f_{\mathrm{nadir}}$ obtained via brute-force method and CE method are same. However, the computation time of CE method is much smaller (the run-time is 45 minutes). We also found that for $n_{\mathbf{S}}=10$, the solution obtained in the third scenario attains a high $f_{\mathrm{nadir}}$ at a low time, whereas for the same number of storage units, the forth scenario shows that it reaches the optimal solution by increasing the number of iterations and random solutions per iteration. The capacity of the storage units considered in the forth scenario is half of the storage capacity considered in the first one, which implies $\{\mathbf{7},\mathsf{4}\}$, $\{\mathbf{10},\mathsf{6}\}$ when $D_{\mathbf{S},\mathrm{total}}^{-1} = 48$ instead of $\{\mathbf{7},\mathsf{2}\}$, $\{\mathbf{10},\mathsf{3}\}$ when $D_{\mathbf{S},\mathrm{total}}^{-1} = 96$. Note that, apart from the first scenario, all the other scenarios have also been examined using he brute-force method, however, they failed to converge to the optimal solutions due to high computational complexity in accordance to \eqref{ratio}.

\begin{table}[htbp]
\centering
\renewcommand{\arraystretch}{1.1}
\caption{Storage distribution and $f_{\mathrm{nadir}}$ considering transients at all renewable resources simultaneously.}
\label{table:IsraelStorageDistAlg2}
\begin{tabular}{c | c c c} 
\toprule
Scenario & $n_{\mathbf{S}}$ & $f_{\mathrm{nadir}}$ [in Hz] & $\left\{\mathbf{Bus},\mathsf{Storage}\right\}$ \\
\midrule
1 & $5$ & $49.4336$ & $\{\mathbf{7},\mathsf{2}\}$, $\{\mathbf{10},\mathsf{3}\}$ \\ 
2 & $8$ & $49.4328$ & $\{\mathbf{7},\mathsf{3}\}$, $\{\mathbf{10},\mathsf{5}\}$ \\ 
3 & $10$ & $49.4256$ & $\{\mathbf{7},\mathsf{3}\}$, $\{\mathbf{10},\mathsf{7}\}$ \\ 
4 & $10$ & $49.4336$ & $\{\mathbf{7},\mathsf{4}\}$, $\{\mathbf{10},\mathsf{6}\}$ \\ 
\bottomrule
\end{tabular}
\end{table}

\begin{figure}[htbp]
\centering
\includegraphics[width=0.7\textwidth]{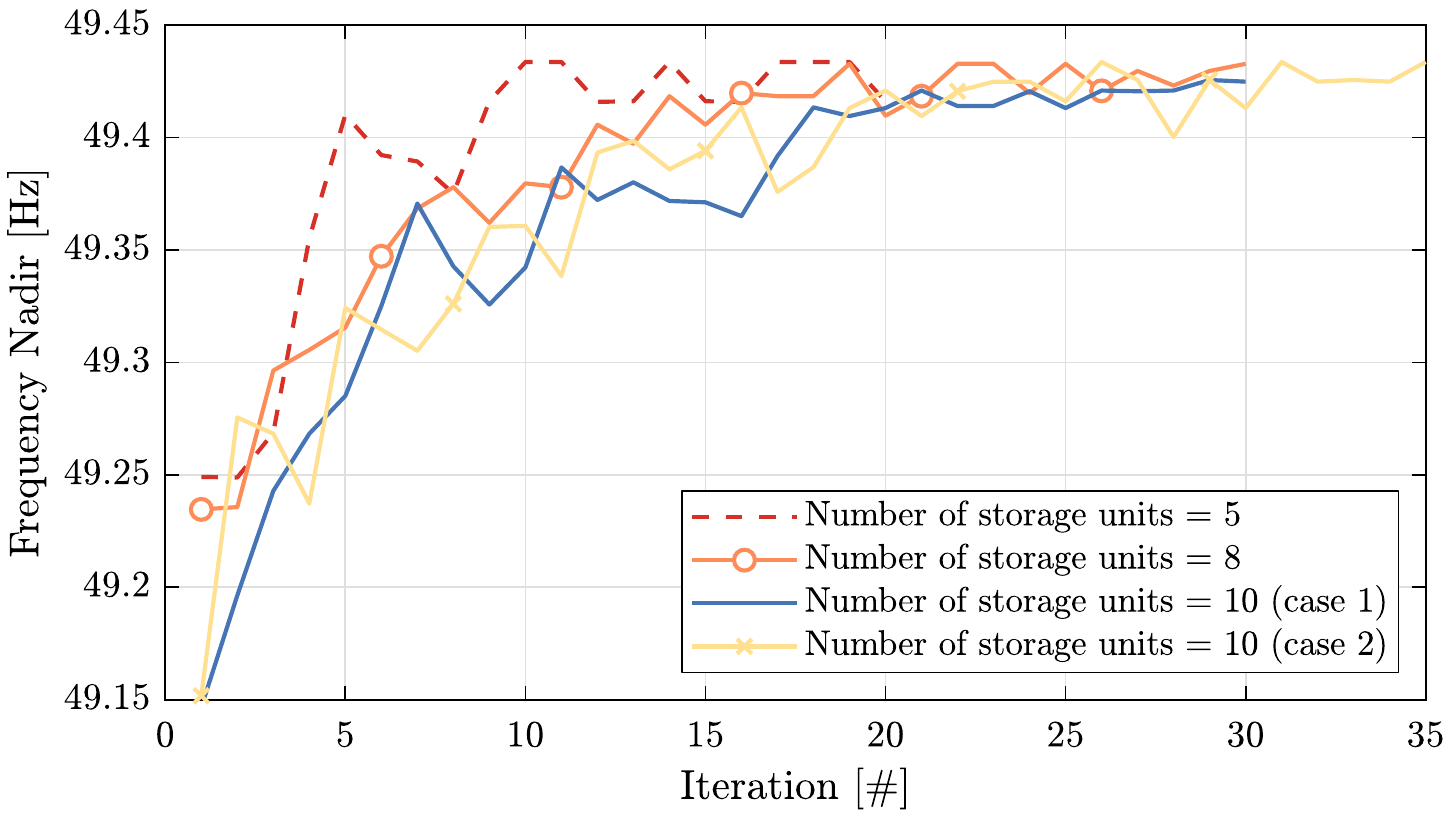}
\caption{ Best $f_{\mathrm{nadir}}$ solution per iteration. }
\label{fig:iterOptPlot}
\end{figure}

For each scenarios, Fig.~\ref{fig:iterOptPlot} illustrates the best solution per iteration. We found that for $n_{\mathbf{S}}=5,8$ and 10 the optimal solutions are obtained at 10, 19, 29 (for Scenario 3) and 26 (for Scenario 4) iteration, respectively.

In Table~\ref{table:PmetricUpdates}, the buses with highest probability to place a storage unit are presented from the probability metric $Q$. Note that for all the above scenarios, only four buses are relevant with total probability higher than 0.7 and all other thirteen buses have total probability lower than 0.3 to have a storage unit. It aligns with the brute-force search for $n_{\mathbf{S}}=5$ and the six best distributions in Table~\ref{table:IsraelStorageDist} which suggest same buses for storage units locations. Furthermore, for the forth scenario which the number of iteration is high, the probability to place a storage in these four relevant buses is higher than 0.8 and the probability to place a storage in one of the two most relevant buses (i.e., bus 7 or 10) is more than 0.64. From these probabilities it can be concluded that the best options to locate the storage systems are buses near or at the buses where the transient occurs.

\begin{table}[H]
\centering
\renewcommand{\arraystretch}{1.1}
\caption{Probability metric at the end of the CE method search.}
\label{table:PmetricUpdates}
\begin{tabular}{c | c c c c c} 
\toprule
Scenario & $n_{\mathbf{S}}$ & $q_7$ & $q_{10}$ & $q_{13}$ & $q_{19}$ \\
\midrule
$1$ & $5$ & $0.1447$ & $0.3899$ & $0.0933$ & $0.0772$ \\
$2$ & $8$ & $0.1617$ & $0.459$ & $0.0907$ & $0.0735$ \\
$3$ & $10$ & $0.143$ & $0.438$ & $0.102$ & $0.0821$ \\
$4$ & $10$ & $0.1808$ & $0.4597$ & $0.1078$ & $0.0635$ \\
\bottomrule
\end{tabular}
\end{table}

As conclusion, we comment that the CE method is simple and provides a solution with low-complexity that can reach to the near-optimal solutions for high dimensional search space. As a result, it allows to rapidly analyze and explore complex planning problems and open academic research questions related to location and size of multiple storage units.

A comparison between the adaptation of  CE method and the brute-force search is presented in Table~\ref{tbl:Concl}. This table may help to understand under which conditions the suggested methods are the most efficient.

\begin{table}[H]
\centering
\renewcommand{\arraystretch}{1.2}
\caption{Brute-force compared to Cross Entropy: advantages and challenges.}
\label{tbl:Concl}
\begin{center}
\begin{tabular}{P{2.5cm} | P{4.5cm} P{3.8cm} l} 
\toprule
Method & Advantages & Challenges & When to use \\
\midrule
Brute-force method & Easy to implement, and  converges to the global optimum & High numeric complexity & complexity ratio \eqref{ratio} $\approx 1$ \\
Cross Entropy method & Low numeric complexity, easy to implement, and near-optimal solutions & Convergence to the global optimal solution is not guaranteed & $\text{complexity ratio}~\eqref{ratio} \gg 1$ \\    
\bottomrule
\end{tabular}
\end{center}
\end{table}

\section{Conclusion} \label{sec:sum}
In recent years the share of renewable sources is increasing and the inertia within the grid is slowly being reduced. In order to achieve better inertial response and frequencies regulation in large power systems, the need for multiple storage units appropriately sized and located is essential. This work formulate the inertial response for the maximum frequency deviation as the main objective when the frequency varies across the network. In this study two numeric approaches are developed based on combinatorial optimization which allow to answer the question of ``how to distribute constant number of storage units in the grid under transient events such that the inertial response of the maximum frequency deviation is minimized?''. The work suggest a time-varying phasor model with energy distributed storage devices connected to the grid using grid-supporting inverters based on droop control mechanism. In this model the total storage capacity is bounded based on the allowed steady-state frequency deviation after disturbances. Two numeric approaches are developed using the suggested model and examined on a case study of the future Israeli grid. While the first approach named brute-force search, reaches to global optimal solution, the second approach, an adaptation of the cross-entropy method, has low computational complexity and should be used when the problem consists of high dimensional solutions. A comprehensive analysis accompanied by comparison to a model which not consider spatial effects is presented. it has been shown that when the frequency is uniform across the grid the inertial response is less accurate since during the transient event the generators' frequencies are not equal.

Our numeric results conclude that the model expectation regarding size and location are aligned to conclusions of other state-of-the-art works- the location of the storage should be in areas of low inertia and/or at the site of disturbances. Furthermore, this work also conclude that the storage units should be placed around the area of the disturbances, including in sites with high inertia in accordance to the network topology. For example storage should be added to a site that generates almost $9\%$ of the entire power in the grid which been effected by a nearby disturbance since it is a central bus in the network. 

Accordingly, based all mentioned above, the suggested approaches should provide guidelines for choosing the best locations and size of distributed storage units for frequency stability, specifically during inertia response. 

As part of future researches, we consider extensions these approaches and model to more complex planning problems and open questions regarding location and size of storage devices for frequency stability. 



\appendix
\section{Determination of $\mathcal{G}$ and $\mathcal{H}$ matrices}\label{ap:FromY2deltaW}
\begin{description}

\item[Step 1] Calculate the admittance matrix $I =YV$ using the DC power flow
\begin{equation}\label{ApDCPF1}
P_i = \sum_{k \neq i} |V|^2 |y_{i,k}|(\delta_i - \delta_k).
\end{equation}	
\item[Step 2] Swap $Y$ matrix such that the order of the buses is generators, storage units and loads. $Y_{\mathrm{sort}} =FYF^{-1}$.
\item[Step 3] calculate the power vector $P$ based on 
\begin{equation}\label{ApDCPF2}
P = |V|^2 \Upsilon {\delta},
\end{equation}	
where
\begin{equation}\label{YbusAdMatrix}
\Upsilon = j\begin{bmatrix}
\sum y_{1k} & -y_{12} & -y_{13} & \dots & -y_{1b} \\
-y_{21} & \sum y_{2k} & -y_{23} & \dots & -y_{2b} \\
-y_{31} & -y_{32} & \sum y_{3k} & \dots & -y_{3b} \\
\vdots& \vdots & \vdots& \ddots & \vdots \\
-y_{b1} & -y_{b2} & -y_{b3} & \dots & \sum y_{bb} \\
\end{bmatrix}, 
\end{equation}	
and $y_{i,k}$ are the variables in $Y_{\mathrm{sort}}$.
\item[Step 4] Find the $W$ and $\Lambda$ from $\Upsilon$ matrix in \eqref{ApDCPF2}.
\begin{equation}\label{ApY2LW}
\begin{bmatrix} P_{gs} \\ P_{L} \end{bmatrix} =
\begin{bmatrix}
U_{11} & U_{12} \\
U_{21} & U_{22} \\
\end{bmatrix}
\begin{bmatrix} \delta_{gs} \\ \delta_{L} \end{bmatrix} ,\\ 
\end{equation}
where $P_{gs}$ is related to the generators and storage power in the system and $P_L$ is related to all power loads and renewable sources.
Thus
\begin{equation}\label{ApDCPF3}
P_{gs} = (U_{11} - U_{12}U_{22}^{-1}U_{21}) \delta_{gs} + U_{12}U_{22}^{-1}P_L,
\end{equation}	
where $U_{22}$ is assumed to be invertible and based on \eqref{dc1} 
\begin{equation}\label{eqWLamda}
\begin{aligned}
& \mathcal{G} = U_{12}U_{22}^{-1}, \\
& \mathcal{H} = (U_{12}U_{22}^{-1}U_{21})) \begin{bmatrix}\mathbf{0} \\ I_{(n_{\mathbf{G}}+n_{\mathbf{S}}-1)} \end{bmatrix} .
\end{aligned}
\end{equation}	
\end{description}

\section{Default values} \label{ap:Consts}
\begin{table}[H]
\centering
\renewcommand{\arraystretch}{1.1}
\begin{tabular}{c l c c}			
\toprule
Constant & Description & Value & Units \\
\midrule
$p_f$ & number of magnetic poles on the rotor & 2 & -- \\
$\omega_0$ & nominal grid frequency & $2 \pi 50$ & [rad/s] \\
$P_{rt}$ & the generator rated power & generator maximum power & [W] \\
$H$ & Inertia constant & 6 & [s] \\
$J$ & rotor moment of inertia & $\frac{2H P_{rt}}{\omega_0^2} (\frac{p_f}{2})^2$ & $[\text{W}\cdot s^3]$ \\
$K$ & swing equation constant & $\frac{1}{J \omega_0^2}(\frac{p_f}{2})^2$ & $[1/(\text{W}\cdot s)]$ \\
$\alpha$ & droop percentage & 0.05 & -- \\
$\alpha_s$ & storage droop percentage & 0.1 & -- \\
$D$ & generator droop-control damping factor & $\alpha \frac{\omega_0}{P_{rt}}$ & $[1/(\text{W}\cdot s)]$ \\
$D_s$ & storage droop-control damping factor & $0 <D_s < 1$ & $[1/(\text{W}\cdot s)]$ \\
\bottomrule
\end{tabular}
\end{table}

\section{Proof of Claim~\ref{claim-1}} \label{ap:FirstBound}
In this section we prove Claim~\ref{claim-1}. This proof relies on several standard assumptions stated below:

\begin{assump} \label{assume1}
In the subsequent analysis, we assume
\begin{enumerate}
\item The power network is based on the DC power flow: the transmission network is balanced three-phase, lossless, and can deliver unlimited power. Furthermore all the generators are lossless. 
\item The mechanical power of each generator denoted as $P_{\mathbf{G},i}^{\mathrm{mech}}(\cdot)\in\mathbb{R}$, is governed by a droop control mechanism as $P_{\mathbf{G},i}^{\mathrm{mech}}(t)=3P_{\mathbf{G},i}^{\mathrm{ref}}(t)-\frac{1}{D_{\mathbf{G},i}}(\omega_{\mathbf{G},i}(t)-\omega_{0})$ for all $i\in\mathcal{N}_{\mathbf{G}}$.
\item The rotor pole $p_{f,i}=2$ for all $i\in\mathcal{N}_{\mathbf{G}}$ thus, $K_i = \frac{1}{J_{i}\omega_{0}}$.
\item The reference power of each storage device $P_{\mathbf{S},i}^{\mathrm{ref}}(t)=0$ for all $i\in\mathcal{N}_{\mathbf{S}}$.
\item There is an inverse proportion between the moment of inertia $J_{i}$ and the generators' constants $D_{\mathbf{G},i}$, such that
\begin{align*}
J_{1}D_{\mathbf{G},1}=J_{2}D_{\mathbf{G},2}=\cdots=J_{n_{\mathbf{G}}} D_{\mathbf{G},n_{\mathbf{G}}}.
\end{align*}
\item At steady-state, all the frequencies are equal, i.e.,
\begin{equation}
\bar{\omega}_{\mathbf{G},1}=\cdots=\bar{\omega}_{\mathbf{G},n_\mathbf{G}}=\bar{\omega}_{\mathrm{coi}}=:\omega_{ss}.
\end{equation}
Here, $\omega_{\mathrm{coi}}(\cdot)\in\mathbb{R}$ is the central frequency of the system, and it is defined as
\begin{equation}\label{coi}
\omega_{\mathrm{coi}}(t):=\frac{1}{J_{\mathrm{tot}}}\sum_{i=1}^{n_{\mathbf{G}}}J_{i}\omega_{\mathbf{G},i}(t),
\end{equation}
where $J_{\mathrm{tot}}:=\sum\limits_{i=1}^{n_{\mathbf{G}}}J_{i}$.
\end{enumerate}
\end{assump}

The formal proof is stated below:
\begin{proof}
Let us recall that the generator dynamics~\eqref{sg2} which is modeled via swing equation and droop control mechanism, is stated below
\begin{equation}
\frac{\mathrm{d}}{\mathrm{d}t}\omega_{\mathbf{G},i}(t)=K_{i}\left(3P_{\mathbf{G},i}^{\mathrm{ref}}(t)-3P_{\mathbf{G},i}(t)-\frac{1}{D_{\mathbf{G},i}}\left(\omega_{\mathbf{G},i}(t)-\omega_0\right)\right) \quad \forall i\in \mathcal{N}_{\mathbf{G}} \label{eq:swing_droop}.
\end{equation} 
First multiplying both sides of~\eqref{eq:swing_droop} by $J_i$ we obtain
\begin{equation}
J_i\frac{\mathrm{d}}{\mathrm{d}t}\omega_{\mathbf{G},i}(t)=-\frac{1}{\omega_0D_{\mathbf{G},i}}\omega_{\mathbf{G},i}(t)+\frac{3}{\omega_{0}}P_{\mathbf{G},i}^{\mathrm{ref}}(t)+\frac{1}{D_{\mathbf{G},i}}-\frac{3}{\omega_{0}}P_{\mathbf{G},i}(t), \label{eq:swing_droop2}
\end{equation}
then combine all the generator equations given in~\eqref{eq:swing_droop2}, we find
\begin{align}
\frac{1}{J_{\mathrm{tot}}}\frac{\mathrm{d}}{\mathrm{d}t}\sum_{i=1}^{n_{\mathbf{G}}} J_{i}\omega_{\mathbf{G},i}(t)&=-\frac{1}{\omega_{0}J_{\mathrm{tot}}}\sum_{i=1}^{n_{\mathbf{G}}}\frac{\omega_{\mathbf{G},i}(t)}{D_{\mathbf{G},i}}+\frac{3}{\omega_{0}J_{\mathrm{tot}}}\sum_{i=1}^{n_{\mathbf{G}}}P_{\mathbf{G},i}^{\mathrm{ref}}(t) \nonumber \\
&+\frac{1}{J_{\mathrm{tot}}}\sum_{i=1}^{n_{\mathbf{G}}}\frac{1}{D_{\mathbf{G},i}}-\frac{3}{\omega_{0}J_{\mathrm{tot}}}\sum_{i=1}^{n_{\mathbf{G}}}P_{\mathbf{G},i}(t), \label{pf1}
\end{align}
where $J_{\text{tot}}$ is defined in Assumption~\ref{assume1}. Since the transmission network is balanced three-phase, lossless, and can deliver unlimited power as stated in Assumption~\ref{assume1}, the total load power $P_{\mathbf{L},\mathrm{tot}}(t)$ can be calculated as
\begin{equation}
P_{\mathbf{L},\mathrm{tot}}(t)=\sum_{i=1}^{n_{\mathbf{L}}}P_{\mathbf{L},i}(t)=\sum_{i=1}^{n_{\mathbf{G}}}P_{\mathbf{G},i}(t). \label{pf2}
\end{equation}
In addition, employing Assumption~\ref{assume1} and the definition of $J_{\text{tot}}$, we can further establish the following relationships
\begin{align}
\sum_{i=1}^{n_{\mathbf{G}}}\frac{\omega_{\mathbf{G},i}(t)}{D_{\mathbf{G},i}}&=\sum_{i=1}^{n_{\mathbf{G}}}\frac{J_{i}\omega_{\mathbf{G},i}(t)}{J_{i}D_{\mathbf{G},i}}=\frac{1}{J_{1}D_{\mathbf{G},i}}\sum_{i=1}^{n_{\mathbf{G}}}J_{i}\omega_{\mathbf{G},i}(t), \nonumber \\
\sum_{i=1}^{n_{\mathbf{G}}}\frac{1}{D_{\mathbf{G},i}}&=\sum_{i=1}^{n_{\mathbf{G}}}\frac{J_i}{J_iD_{\mathbf{G},i}}=\frac{J_{\mathrm{tot}}}{J_1 D_{\mathbf{G},1}}. \label{pf3}
\end{align}
Let $P_{\text{tot}}^{\mathrm{ref}}(\cdot)\in\mathbb{R}$ be the total reference power, and it can be calculated as $P_{\text{tot}}^{\mathrm{ref}}(t)=\sum_{i=1}^{n_{\mathbf{G}}}P_{\mathbf{G},i}^{\mathrm{ref}}(t)+\sum_{i=1}^{n_{\mathbf{S}}}P_{\mathbf{S},i}^{\mathrm{ref}}(t)$. Since $P_{\mathbf{S},i}^{\mathrm{ref}}(t)=0$ for all $i\in\mathcal{N}_{\mathbf{S}}$ as stated in Assumption~\ref{assume1}, $P_{\mathrm{tot}}^{\mathrm{ref}}(t)$ reduces to
\begin{equation}
P_{\text{tot}}^{\mathrm{ref}}(t)=\sum_{i=1}^{n_{\mathbf{G}}}P_{\mathbf{G},i}^{\mathrm{ref}}(t). \label{pf4}
\end{equation}
Now substituting \eqref{pf2}, \eqref{pf3} and~\eqref{pf4} in~\eqref{pf1}, and using the definition of $\omega_{\text{coi}}(t)$ given in~\eqref{coi}, we obtain
\begin{align}
\frac{\mathrm{d}}{\mathrm{d}t}\omega_{\mathrm{coi}}(t)&=-\frac{1}{\omega_{0}J_1D_{\mathbf{G},1}}\left(\omega_{\mathrm{coi}}(t)-\omega_0\right)
+\frac{3}{\omega_{0}J_{\mathrm{tot}}}\left(P_{\mathrm{tot}}^{\mathrm{ref}}(t)-P_{\mathbf{L},\mathrm{tot}}(t)\right). \label{pf5}
\end{align}
Let us define $\Delta\omega(t):=\omega_{\mathrm{coi}}(t)-\omega_{0}$ and $\Delta P_{\mathbf{L}}(t):=P_{\mathbf{L},\mathrm{tot}}(t)-P_{\mathrm{tot}}^{\mathrm{ref}}(t)$. Then, employing these definitions, Eq.~\eqref{pf5} can further be simplified as
\begin{equation}\label{pf6}
\frac{\mathrm{d}}{\mathrm{d}t}\Delta\omega(t)=-\frac{1}{\omega_{0}J_1 D_{\mathbf{G},1}}\Delta\omega(t)-\frac{3}{\omega_{0}J_{\mathrm{tot}}}\Delta P_{\mathbf{L}}(t),
\end{equation}
which can be denoted as the \textit{aggregated swing equation}. Next we intend to evaluate the steady-state frequency deviation $\Delta\omega_{ss}(\cdot)\in\mathbb{R}$. Since at steady-state all the frequencies are same as stated in Assumption~\ref{assume1}, we substitute $\frac{\mathrm{d}}{\mathrm{d}t}\Delta\omega(t)=0$ in~\eqref{pf6}, which leads to
\begin{equation}
\Delta\omega_{ss}(t)=\frac{3J_1D_{\mathbf{G},1}}{J_{\mathrm{tot}}}\Delta P_{\mathbf{L}}(t), \label{pf7}
\end{equation} 
and it is further represented employing the relationships in~\eqref{pf3} as
\begin{equation}
\Delta \omega_{ss}(t)=\frac{3}{\sum\limits_{i=1}^{n_{\mathbf{G}}}\frac{1}{D_{\mathbf{G},i}}}\Delta P_{\mathbf{L}}(t). \label{pf8}
\end{equation}
Let $P_{\mathbf{S},\mathrm{tot}}(\cdot)\in\mathbb{R}$ be the total power of the storage devices, and it is calculated as $P_{\mathbf{S},\mathrm{tot}} = \sum_{i=1}^{n_{\mathbf{S}}} P_{\mathbf{S},i}(t)$. Now we assume that the deviation of the load power $\Delta P_{\mathbf{L}}(\cdot)\in\mathbb{R}$ is solely caused by the power transients, which leads to $\Delta P_{\mathbf{L}}(t)=P_{\mathbf{S},\mathrm{tot}}(t)-\mathsf{P}_{\mathrm{trans}}(t)$. Based on this assumption, \eqref{pf8} can be written as
\begin{equation}
\Delta\omega_{ss}=\frac{3}{\sum\limits_{i=1}^{n_{\mathbf{G}}}\frac{1}{D_{\mathbf{G},i}}}\left(P_{\mathbf{S},\mathrm{tot}}(t)-\mathsf{P}_{\mathrm{trans}}(t)\right). \label{pf200}
\end{equation}
Revisiting~\eqref{sd1}, we obtain the steady-state power of the $i^{th}$ storage device where $i\in\mathcal{N}_{\mathbf{S}}$, as
\begin{equation}
P_{\mathbf{S},i}(t)=-\frac{\omega_{\mathbf{S},i}(t)-\omega_{0}}{3 D_{\mathbf{S},i}}=-\frac{\omega_{\mathrm{coi}}- \omega_{0}}{3D_{\mathbf{S},i}}=-\frac{\Delta \omega_{ss}}{3D_{\mathbf{S},i}},
\end{equation}
which further leads to
\begin{equation}
P_{\mathbf{S},\mathrm{tot}}(t)=-\frac{\Delta\omega_{ss}}{3}\left(\sum_{i=1}^{n_{\mathbf{S}}}\frac{1}{D_{\mathbf{S},i}}\right).\label{pf100} 
\end{equation}
Substituting~\eqref{pf100} in~\eqref{pf200} we find 
\begin{align*}
\sum_{i=1}^{n_{\mathbf{S}}}\left(\frac{1}{D_{\mathbf{S},i}}\right)=\frac{3\mathsf{P}_{\mathrm{trans}}(t)}{\Delta\omega_{ss}}-\sum_{i=1}^{n_{\mathbf{G}}}\left(\frac{1}{D_{\mathbf{G},i}}\right).
\end{align*}
Therefore, in order to keep a bounded steady-state frequency, the total damping coefficient of the storage devices needs to obey 
\begin{align*}
\sum_{i=1}^{n_{\mathbf{S}}}\frac{1}{D_{\mathbf{S},i}}\geq\frac{3\mathsf{P}_{\mathrm{trans}}(t)}{\Delta \omega_{ss,\max}}- \sum_{i=1}^{n_{\mathbf{G}}}\frac{1}{D_{\mathbf{G},i}}, 
\end{align*}
which verifies our claim.
\end{proof}

\bibliographystyle{unsrtnat}
\bibliography{template}

\end{document}